\documentclass[runningheads,openany]{svmult}

\usepackage{makeidx}   % allows index generation
\usepackage{graphicx}  % standard LaTeX graphics tool
\usepackage{subeqnar}  % subnumbers individual equations
\usepackage{multicol}  % used for the two-column index
\usepackage{physprbb}    % centered layout of diverse elements, etc.
\makeindex             % used for the subject index
\usepackage{amsmath}
\usepackage{amsfonts}
\usepackage{theorem}

\usepackage{epsfig}

\newcommand{\enquote}[1]{'#1'}
\newcommand{\paverage}[1]{\left\langle #1 \right\rangle_{\rm P}}
\newcommand{\cov}{{\ifmmode{\text{{\it cov}}}\else{ {\it cov} }\fi}}
\newcommand{\cor}{{\ifmmode{\text{{\it cor}}}\else{ {\it cor} }\fi}}
\newcommand{\var}{{\ifmmode{\text{{\it var}}}\else{ {\it var} }\fi}}

\begin{document}

\title{Mark correlations: relating physical properties to spatial
distributions} 
\author{Claus Beisbart$^1$, Martin Kerscher$^2$, Klaus
Mecke$^{3,4}$} 
\authorrunning{Beisbart, Kerscher, Mecke}

\institute{
University of Oxford, 
Nuclear \& Astrophysics Laboratory, Keble Road, Oxford OX1 3RH, Great Britain
\and
Ludwig-Maximilians-Universit\"{a}t,  Sektion
Physik,  Theresienstra{\ss}e 37,  D-80333  M\"{u}nchen, Germany
\and
Max-Planck-Institut f\"ur Metallforschung, Heisenbergstr. 1, D-70569
Stuttgart, Germany   
\and
Institut f\"ur Theoretische und Angewandte Physik, Fakult\"at f\"ur
Physik, Univer\-si\-t\"at Stuttgart, Pfaffenwaldring 57, D-70569 Stuttgart, 
Germany
}
\date{draft \today}

\maketitle
draft \today
%%%%%%%%%%%
\begin{abstract}
Mark  correlations provide a  systematic approach  to look  at objects
both  distributed  in space  and  bearing  intrinsic information,  for
instance  on  physical  properties.   The interplay  of  the  objects'
properties (marks)  with the spatial  clustering is of  vivid interest
for many applications; are, e.g., galaxies with high luminosities more
strongly clustered than dim ones?   Do neighbored pores in a sandstone
have similar sizes?  How does the  shape of impact craters on a planet
depend on the geological surface properties?  In this article, we give
an introduction  into the  appropriate mathematical framework  to deal
with  such questions,  i.e.   the theory  of  marked point  processes.
After having  clarified the notion  of segregation effects,  we define
universal test quantities applicable to realizations of a marked point
processes.  We show their power  using concrete data sets in analyzing
the luminosity-dependence  of the galaxy clustering,  the alignment of
dark   matter  halos  in   gravitational  $N$-body   simulations,  the
morphology- and diameter-dependence of the Martian crater distribution
and the size correlations of pores in sandstone.
In order to understand our data in more detail, we discuss the Boolean
depletion  model, the  random field  model  and the  Cox random  field
model.    The  first   model  describes   depletion  effects   in  the
distribution of  Martian craters and  pores in sandstone,  whereas the
last   one  accounts   at   least  qualitatively   for  the   observed
luminosity-dependence of the galaxy clustering.
\end{abstract}

%%%
\section{Marked point sets}
\label{sec:kerscher_basic}
\begin{figure}
\begin{center}
\epsfig{file=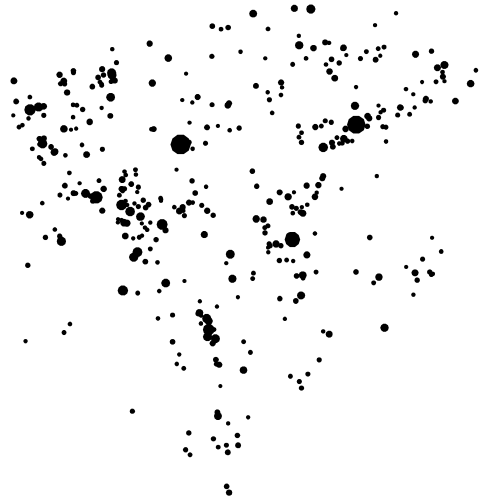,width=8cm}
\end{center}
\caption{The  galaxy  distribution  as  traced  by  the  Southern  Sky
Redshift Survey 2 (SSRS~2). We show a part of the sample investigated,
projected down into two  dimensions.  Each circle represents a galaxy,
its  radius is proportional  to the  galaxy's luminosity.  For further
details see Section~\ref{sec:kerscher_gal}.\label{fig:kerscher_galaxies-circles}}
\end{figure}

Observations of  spatial patterns at various  length scales frequently
are the only point where  the physical world meets theoretical models.
In many cases these patterns consist of a number of comparable objects
distributed in space  such as pores in a sandstone,  or craters on the
surface    of   a    planet.     Another   example    is   given    in
Figure~\ref{fig:kerscher_galaxies-circles},   where  we   display   the  galaxy
distribution as traced by a  recent galaxy catalogue. The galaxies are
represented as  circles centered at their positions,  whereas the size
of the circles mirrors the luminosity of a galaxy. In order to test to
which  extent  theoretical   predictions  fit  the  empirically  found
structures  of that  type, one  has to  rely on  quantitative measures
describing the physical  information.  Since theoretical models mostly
do not try to explain  the structures individually, but rather predict
some of their generic properties,  one has to adopt a {\em statistical
point of view} and to interpret  the data as a realization of a random
process.  In  a first step one  often confines oneself  to the spatial
distribution of the objects constituting the patterns and investigates
their clustering  thereby thinking  of it as  a realization of  a {\em
point process}.   Assuming that  perspective, however, one  neglects a
possible  linkage between  the  spatial clustering  and the  intrinsic
properties of the objects.  For instance, there are strong indications
that the clustering of galaxies depends on their luminosity as well as
on        their         morphological        type.         Considering
Figure~\ref{fig:kerscher_galaxies-circles},  one   might  infer  that  luminous
galaxies are more strongly correlated than dim ones. Effects like that
are referred to as {\em mark segregation} and provide insight into the
generation and interactions of,  e.g., galaxies or other objects under
consideration. The  appropriate statistical framework  to describe the
relation between the spatial distribution of
physical objects and their inner properties are {\em marked point
processes}, where discrete, scalar-, or vector-valued marks are
attached to the random points.\\
In this contribution we outline how to describe marked point
processes; along that line we discuss two notions of independence
(Section~\ref{sec:kerscher_basic}) and define corresponding statistics that
allow us to quantify possible dependencies.  After having shown that
some empirical data sets show significant signals of mark segregation
(Section{}\ref{sec:kerscher_data}), we turn to analytical models, both
motivated by mathematical and physical considerations
(Section~{}\ref{sec:kerscher_models}).
\\
Contact  distribution functions  as presented  in the  contribution by
D.~Hug et al.  in this volume are an  alternative technique to measure
and  statistically quantify distances  which finally  can be  used to
relate  physical properties  to spatial  structures.  Mark correlation
functions  are useful  to  quantify molecular  orientations in  liquid
crystals (see  the contribution by  F.~Schmid and N.~H.~Phuong  in this
volume)   or   in  self-assembling   amphiphilic   systems  (see   the
contribution by U.~S.~Schwarz and G.~Gompper in this volume). But also
to  study anisotropies  in composite  or porous  materials,  which are
essential for elastic and transport properties (see the contributions
by D.~Jeulin,  C.~Arns et  al. and H.-J.~Vogel  in this  volume), mark
correlations may be relevant.
%%%%%%%%%%%%%%%%%%%%%%%%%%%%%%%%%%%%%%%%%%%%%%%%%%
%  basic 
%%%%%%%%%%%%%%%%%%%%%%%%%%%%%%%%%%%%%%%%%%%%%%%%%%
\subsection{The framework}
The empirical data -- the positions ${\mathbf{x}}_i$ of some objects together
with their intrinsic properties $m_i$ -- are interpreted as a
realization of a marked point process $\{({\mathbf{x}}_i,m_i)\}_{i=1}^N$
 (Stoyan, Kendall and Mecke, 1995).
For simplicity we restrict ourselves to homogeneous and isotropic
processes.
\\
The hierarchy of joint probability densities provides a suitable tool
to describe the stochastic properties of a marked point process.
Thus, let $\varrho_1^{{\cal S}{\cal M}}\left(({\mathbf{x}},m)\right)$ denote the probability
density of finding a point at ${\mathbf{x}}$ with a mark $m$. For a homogeneous
process this splits into
$\varrho_1^{{\cal S}{\cal M}}\left(({\mathbf{x}},m)\right)=\varrho{\cal M}_1(m)$ where
$\varrho$ denotes the mean number density of points in space
and ${\cal M}_1(m)$ is the probability density of finding the mark $m$ on
an arbitrary point. Later on we need moments of this mark
distribution;  for real-valued marks the $k$th-moment of the
mark-distribution is defined as
\begin{equation}
\overline{m^k} = \int{\rm d} m\ {\cal M}_1(m) m^k;
\end{equation}
the mark variance is $\sigma_M^2=\overline{m^2}-\overline{m}^2$.
\\
Accordingly,   $\varrho_2^{{\cal S}{\cal M}}\left(({\mathbf{x}}_1,m_1),({\mathbf{x}}_2,m_2)\right)$
quantifies the probability  density to find two points  at ${\mathbf{x}}_1$ and
${\mathbf{x}}_2$  with marks  $m_1$ and  $m_2$, respectively  (for second-order
theory        of        marked        point       processes        see
{}\cite{kerscher_stoyan:stochgeom,kerscher_stoyan:fractals}).  It effectively depends
only on $m_1$, $m_2$, and  the pair separation $r=|{\mathbf{x}}_2-{\mathbf{x}}_1|$ for a
homogeneous and isotropic  process. Two-point properties certainly are
the simplest non-trivial  quantities for homogeneous random processes,
but it may be necessary to  move on to higher correlations in order to
discriminate between certain models.

%%%%%%%%%%%%%%%%%%%%%%%%%%%%%%%%%%%%%%%%%%%%%%%%%%
%  Two notions 
%%%%%%%%%%%%%%%%%%%%%%%%%%%%%%%%%%%%%%%%%%%%%%%%%%
\subsection{Two notions of independence}
In the following we will discuss two notions of independence, which
may arise for marked point patterns. For this, consider two Renaissance
families, call them the Sforza and the Gonzaga.  They used to build
castles spread out more or less homogeneously over Italy. In order to
describe this example in terms of a marked point process, we consider
the locations of the castles as points on a map of Italy, and treat a
castle's owner as a discrete mark, $S$ and $G$, respectively.  There
are many ways how the castles can be built and related to each other.

\paragraph{Independent sub-point processes:}
 For example,  the Sforza  may build their  castles regardless  of the
Gonzaga  castles. In  that case  the probability  of finding  a Sforza
castle at ${\mathbf{x}}_1$ and a  Gonzaga castle at ${\mathbf{x}}_2$ factorizes into two
one-point  probabilities and  we  can  think of  the  Sforza and  the
Gonzaga castles as uncorrelated  sub-point processes. In the language
of marked point processes this means, e.g., that
\begin{gather}  
\begin{split}  
\varrho_2^{{\cal S},{\cal M}}\left(  ({\mathbf{x}}_1,m_1)  ,({\mathbf{x}}_2,m_2)\right) &=
\varrho_1^{{\cal S}{\cal M}}\left(  ({\mathbf{x}}_1,m_1)\right)\varrho_1^{{\cal S}{\cal M}}\left(
({\mathbf{x}}_2,m_2)\right) \\
&= \varrho^2 {\cal M}_1(m_1){\cal M}_1(m_2) ,
\end{split}
\end{gather}
for any $m_1\neq m_2$. If all the joint $n$-point densities factorize
into a product of $n'$-point densities of one type each, then we
speak of {\em independent sub-point processes}.  Dependent sub-point
processes indicate {\em interactions} between points of different
marks; for instance, the Gonzaga may build their castles close to the
Sforza ones in order to avoid that a region becomes dominated by the
other family's castles.

\paragraph{Mark-independent clustering:} A second type of
independence  refers to  the question  whether the  different families
have  different  styles to  plan  their  castles.   For instance,  the
Gonzaga  may distribute  their  castles in  a  grid-like manner  over
Italy, whereas the  Sforza may incline to build  a second castle close
to each  castle they  own. Rather than  asking whether  two sub-point
processes (namely  the Gonzaga  and the Sforza  castles, respectively)
are  independent  (``independent   sub-point  processes''),  we  are  now
discussing  whether  they  are   {\em  different}  as  regards  their
statistical clustering properties. Any such difference means that the
clustering {\em depends}  on the  intrinsic mark of  a point.  
\\
Whenever the
two-point probability  density of finding two objects  at ${\mathbf{x}}_1$ and
${\mathbf{x}}_2$ depends on the objects'  intrinsic properties we speak of {\em
mark-dependent clustering}.  It is  useful to rephrase this statement
by using Bayes' theorem and the conditional mark probability density
\begin{gather}  
\label{eq:kerscher_cond-mark-density}
{\cal M}_2(m_1,m_2|{\mathbf{x}}_1,{\mathbf{x}}_2)=
\frac{\varrho_2^{{\cal S},{\cal M}}\left(({\mathbf{x}}_1,m_1),({\mathbf{x}}_2,m_2)\right)}
{\varrho_2^{{\cal S}}\left({\mathbf{x}}_1,{\mathbf{x}}_2\right)},
\end{gather}
in case the spatial product density $\varrho_2^{{\cal S}}(\cdot)$ does not
vanish.  ${\cal M}_2(m_1,m_2|{\mathbf{x}}_1,{\mathbf{x}}_2)$ is the probability density of
finding the marks $m_1$ and $m_2$ on objects located at ${\mathbf{x}}_1$ and
${\mathbf{x}}_2$, given that there are objects at these points.  Clearly,
${\cal M}_2(m_1,m_2|{\mathbf{x}}_1,{\mathbf{x}}_2)$ depends only on the pair separation
$r=|{\mathbf{x}}_1-{\mathbf{x}}_2|$ for homogeneous and isotropic point processes.  We
speak of {\em mark-independent} clustering, if
${\cal M}_2(m_1,m_2|r)$ factorizes
\begin{equation}  
{\cal M}_2(m_1,m_2|r) = {\cal M}_1(m_1) {\cal M}_1(m_2) 
\end{equation}
and thus does not depend on the pair separation. That means that
regarding their marks, pairs with a separation $r$ are not different
from any other pairs.  On the contrary, mark-dependent clustering or
{\em mark segregation} implies that the marks on certain pairs show
deviations from the global mark distribution.
\\
In order to distinguish between both sorts of independencies, let us
consider the case where we are given a map of Italy only showing the
Gonzaga castles.  If the distribution of castles in Italy can be
understood as consisting of independent sub-point processes, we
cannot infer anything about the Sforza castles from the Gonzaga ones.
However, if $\varrho_2^{{\cal S},{\cal M}}\left(({\mathbf{x}}_1,S),({\mathbf{x}}_2,G)\right)
>\varrho^2 {\cal M}_1(S){\cal M}_1(G)$, Sforza castles are likely to be found
close to Gonzaga ones. Here, ${\cal M}_1(S)$ and ${\cal M}_1(G)$ are the
probabilities that a castle belongs to the Sforza or Gonzaga family.
If, on  the other  hand, mark-independent clustering  applies, typical
clustering  properties such  as  the spatial  clustering strength  are
equal  for both  castle  distributions, and  the  Gonzaga castles  are
in the statistical sense already  representative of  the  whole castle  distribution in  Italy.
That means in  particular that, if the Gonzaga  castles are clustered,
so are the Sforza ones. 

Before we turn to applications, we have to develop practical test quantities in order to test for
segregation effects in real data and to describe them in more detail.

%%%
\subsection{Investigating the independence of sub-point processes}

To  investigate  correlations  between sub-point  processes,  suitably
extended nearest neighbor distribution functions or $K$-functions have
been   employed   {}\cite{kerscher_cox:point,kerscher_diggle:statistical}.   Also   the
(conditional)   cross-correlation   functions   can   be   used   (see
Eq.~\ref{eq:kerscher_cond-crosscorr}), for a further test see
{}\cite{kerscher_stoyan:fractals}, p.~302.
Here we consider a multivariate extension of the $J$-function
{}\cite{kerscher_vanlieshout:j}, as suggested by
{}\cite{kerscher_vanlieshout:indices}.
\\
For this, consider the nearest neighbor's distance distribution from
an object with mark $m_i$ to other objects with mark $m_j$,
$G_{ij}(r)$ (``$i$ to $j$'', for details see
{}\cite{kerscher_vanlieshout:indices}).  Let $G_{i\circ}(r)$ denote the
distribution of the nearest neighbor's distance from an object of
type $i$ to any other object (denoted by $\circ$).  Finally,
$G_{\circ\circ}(r)$ is the nearest neighbor distribution of all
points.  Similar extensions of the empty space function are possible,
too. Let $F_i(r)$ denote the distribution of the nearest $i$-object's
distance from an arbitrary position, whereas $F_\circ(r)$ is the
nearest object's distance distribution from a random point in space to
any object in the sample.  We consider the following quantities:
\begin{gather}
\label{eq:kerscher_def-multivar-J}
\begin{split}  
J_{ij}(r) =\frac{1-G_{ij}(r)}{1-F_j(r)} ,\
J_{i\circ}(r) =\frac{1-G_{i\circ}(r)}{1-F_\circ(r)} ,\
J(r) =\frac{1-G_{\circ\circ}(r)}{1-F_\circ(r)} ,
\end{split}
\end{gather}
They are defined whenever $F_j(r),F_\circ(r) <1$. If two sub-point
processes, defined by marks $i\neq j$, are independent then one
gets {}\cite{kerscher_vanlieshout:indices}
\begin{gather}
\label{eq:kerscher_jijone}
J_{ij}(r) = 1 .
\end{gather}
Note, that the $J_{ij}$ depend on higher-order correlations
functions, similar to the $J$-function
{}\cite{kerscher_kerscher:constructing}.  Suitable estimators for these
$J$-functions are derived from estimators of the $F$ and
$G$-functions {}\cite{kerscher_stoyan:stochgeom,kerscher_baddeley:sampling}.

%%%
\subsection{Investigating mark segregation}
\label{sec:kerscher_mark-segregation}
In order to quantify the  mark-dependent clustering or to look for the
mark  segregation,  it  proves  useful to  integrate  the  conditional
probability density ${\cal M}_2(m_1,m_2|r)$ over the marks weighting with a
test                       function                       $f(m_1,m_2)$
{}\cite{kerscher_stoyan:oncorrelations,kerscher_stoyan:stochgeom}.     This   procedure
reduces the number  of variables and leaves us  with the weighted pair
average:
\begin{equation}  
\label{eq:kerscher_def-paverage}
\paverage{f} = \int{\rm d}{m_1}\int{\rm d}{m_2}\ f(m_1,m_2) {\cal M}_2(m_1,m_2|r) .
\end{equation}
The choice of an appropriate weight-function depends on whether the
marks are non-quantitative labels or continuous physical quantities. 
\begin{enumerate}
\item[1.\ ] For labels only combinations of indicator-functions are
possible,  the integral  degenerates into a  sum over  the labels.
Supposed  the marks  of our  objects belong  to classes  labelled with
$i,j,\ldots$, the conditional cross-correlation functions are given by
\begin{equation}
\label{eq:kerscher_cond-crosscorr}
C_{i j}(r) \equiv
\paverage{\delta_{m_1i} \delta_{m_2j} +
(1-\delta_{ij}) \delta_{m_2i} \delta_{m_1j}} (r),
\end{equation}
with  the Kronecker $\delta_{m_1i}=1$  for $m_1=i$  and zero
otherwise.
Mark           segregation          is           indicated          by
$C_{ij}\ne2\varrho_i\varrho_j/\varrho^2$    for     $i\ne    j$    and
$C_{ii}\ne\varrho_i^2/\varrho^2$, where $\varrho_i$ denotes the number
density of points with  label $i$.  The $C_{ij}$ are cross-correlation
functions under the {\em condition} that two points are separated by a
distance  of~$r$  (compare  {}\cite{kerscher_stoyan:fractals},  p.~264,  for
applications   see  the   Martian  crater   distribution   studied  in
Sect.~\ref{sec:kerscher_martian}    and    Figure~\ref{fig:kerscher_mars-markcorr}    in
particular).
\item[2.\  ] For positive  real-valued marks  $m$, the  following pair
averages     prove     to      be     powerful     and     distinctive
{}\cite{kerscher_schlather:mark,kerscher_beisbart:luminosity}: \\
\noindent
\begin{enumerate}
\item
One of the most simplest weights to be used is the mean mark:
\begin{equation}
k_{m}(r) \equiv \frac{\paverage{m_1+m_2}(r)}{2\ \overline{m}} .
\end{equation}
quantifies the deviation of the mean mark on pairs with separation $r$
from the overall mean mark $\overline{m}$.  A $k_{m}>1$ indicates mark
segregation for point pairs with a separation $r$, specifically their
mean mark is then larger than the overall mark average.\\
%\item 
Closely related is Stoyan's $k_{mm}$ function using the 
squared geometric mean of the marks as a weight 
{}\cite{kerscher_stoyan:oncorrelations,kerscher_stoyan:fractals}
\begin{equation}
k_{mm}(r) \equiv \frac{\paverage{m_1m_2}(r)}{\overline{m}^2} .
\end{equation}
\item 
Accordingly, higher moments of the marks may be used to quantify mark
segregation, like the mark fluctuations
\begin{equation}
\var(r) \equiv \paverage{\left(m_1-\paverage{m_1}(r)\right)^2}(r) ,
\end{equation}
or the mark-variogram {}\cite{kerscher_waelder:variograms,kerscher_stoyan:variograms}:
\begin{equation}
\gamma(r) \equiv \paverage{\tfrac{1}{2}\left(m_1-m_2\right)^2}(r) ,
\end{equation}
\item
The mark covariance {}\cite{kerscher_cressie:statistics} is
\begin{equation}
\label{eq:kerscher_def-cov}
\cov(r) \equiv \paverage{m_1 m_2}(r) -\paverage{m_1}(r)\paverage{m_2}(r) .
\end{equation}
Mark segregation can be detected by looking whether $\cov(r)$ differs
from zero. A $\cov(r)$ larger than zero, e.g., indicates that points
with separation $r$ tend to have similar marks.  Sometimes the mark
covariance is normalized by the fluctuations
{}\cite{kerscher_isham:marked}: $\cov(r)/\var(r)$. 
\end{enumerate}
These conditional mark correlation functions can be calculated from
only three independent pair averages~\cite{kerscher_schlather:mark}: $\paverage{m}(r)$,
$\paverage{m_1m_2}(r)$, and $\paverage{m^2}(r)$. Thus the above
mentioned characteristics are not independent, e.g.\
$\var(r)=\gamma(r)+cov(r)$.
\\
We apply  these mark correlation functions to  the galaxy distribution
in   Section~\ref{sec:kerscher_gal}  (Figure~\ref{fig:kerscher_ssrs-lum}),   to  Martian
craters  in Section~\ref{sec:kerscher_martian} (Figure~\ref{fig:kerscher_mars-markcorr})
and to pores in sandstones considered in Section~\ref{sec:kerscher_sandstone}.
%
%%%
\item[3.\ ]
Also vector-valued information ${\mathbf{l}}_i$, describing, e.g., the 
orientation of an anisotropic object at  position ${\mathbf{x}}_i$ may be available. It is therefore interesting to consider vector marks such as done  by
{}\cite{kerscher_ohser:onsecond,kerscher_penttinen:statistical,kerscher_stoyan:fractals} who use a
mark correlation function to quantify the alignment of vector marks.
Here   we  suggest  three   mark  correlation   functions  quantifying
geometrically different  possibilities of  an alignment.  In  order to
ensure coordinate-independence of our  descriptors, we focus on scalar
combinations of the  vector marks in using the  scalar product $\cdot$
and the  cross product  $\times$.  Different from  the case  of scalar
marks,  it  is  a  non-trivial  task  to find  a  set  of  vector-mark
correlation functions which contain all possible information (at least
up to a  fixed order in mark space).  We  provide a systematic account
of how  to construct suitable  vector-mark correlation functions  in a
complete and unique way for general dimensions in the Appendix. \\
Here we only cite  the most important  results.  For that we need  the
distance vector   between   two  points,  ${\mathbf{r}}\equiv{\mathbf{x}}_1-{\mathbf{x}}_2$,  the
normalized distance vector, $\hat{\mathbf{r}}\equiv{\mathbf{r}}/r$,  and the normalized  vector
mark: $\hat{\mathbf{l}}_i\equiv{\mathbf{l}}_i/l_i$  with $l_i=|{\mathbf{l}}_i|$. The    following
conditional mark   correlation  functions will  be  used  to  quantify
alignment effects:
\begin{enumerate}
\item
${\cal A}(r)$ quantifies the ${\cal A}$lignment of the two vector marks ${\mathbf{l}}_1$ and
${\mathbf{l}}_2$:
\begin{equation}
{\cal A}(r) = \frac{1}{\overline{l}^2} \paverage{{\mathbf{l}}_1\cdot{\mathbf{l}}_2}(r)\;\;\;.
\label{eq:kerscher_align}
\end{equation}
It is proportional to the cosine of the angle between
${\mathbf{l}}_1$ and ${\mathbf{l}}_2$. We normalize  with the mean $\overline{l}$. For purely independent vector marks ${\cal A}(r)$ is zero, whereas ${\cal A}(r)>0$ means that the marks of pairs separated by $r$ tend to align parallel to each other. -- In some applications, e.g. for the orientations of ellipsoidal objects, the vector mark is only defined up to a sign,  i.e.\ ${\mathbf{l}}$ and $-{\mathbf{l}}$ mean actually the same. In this case the absolute value of the scalar product is useful:
\begin{equation}
\label{eq:kerscher_def-vector-corr-absolute}
{\cal A}'(r) \equiv \frac{1}{\overline{l}^2} \paverage{|{\mathbf{l}}_1\cdot{\mathbf{l}}_2|}(r)\;\;\;.
\end{equation}
For uncorrelated random vectors we get ${\cal A}'(r)=1/2$. ${\cal A}$ and ${\cal A}'$
can  readily be  generalized to  any  dimension $d$,  where we  expect
${\cal A}'=$
$\pi^{-\frac{1}{2}}\frac{\Gamma(\frac{d}{2})}{\Gamma(\frac{d+1}{2})}$
for  uncorrelated random  orientations.  In two  dimensions ${\cal A}'$  is
proportional to $k_d$ as defined by~\cite{kerscher_stoyan:fractals}.
\item
${\cal F}(r)$  quantifies  the  ${\cal F}$ilamentary  alignment of  the  vectors
${\mathbf{l}}_1$  and ${\mathbf{l}}_2$  with respect  to the  line connecting  both halo
positions:
\begin{equation}
{\cal F}(r) \equiv \frac{1}{2\ \overline{l}} 
\paverage{|{\mathbf{l}}_1\cdot{\hat{\mathbf{r}}}| + |{\mathbf{l}}_2\cdot{\hat{\mathbf{r}}}|}(r), 
\label{eq:kerscher_filament}
\end{equation}
${\cal F}(r)$ is  proportional to the  cosine of the angle  between ${\mathbf{l}}_1$
and  the  distance  vector   $\hat{\mathbf{r}}$  connecting  the  points.   For
uncorrelated  random  vector  marks,  we  expect  again  ${\cal F}(r)=1/2$;
${\cal F}(r)$ becomes  larger than that,  whenever the vector marks  of the
objects tend  to point to  objects separated by  $r$ -- an  example is
provided  by  rod-like metallic  grains  in  an  electric field:  they
concentrate along  the field lines  and orient themselves  parallel to
the field lines.
\item
${\cal P}(r)$ quantifies the ${\cal P}$lanar alignment of the vectors and the
distance vector. ${\cal P}(r)$ is proportional to the volume of the
rhomb defined by ${\mathbf{l}}_1$, ${\mathbf{l}}_2$ and $\hat{\mathbf{r}}$:
\begin{equation}
{\cal P}(r) = \frac{1}{2\overline{l}^2}\ \paverage{\left|
{\mathbf{l}}_1\cdot\frac{{\mathbf{l}}_2\times{\hat{\mathbf{r}}}}{|\hat{\mathbf{l}}_2\times{\hat{\mathbf{r}}}|} \right|
+ \left|
{\mathbf{l}}_2\cdot\frac{{\mathbf{l}}_1\times{\hat{\mathbf{r}}}}{|\hat{\mathbf{l}}_1\times{\hat{\mathbf{r}}}|} \right|}(r),
\label{eq:kerscher_planar}
\end{equation}
Quite  obviously, this quantity  can not  be generalized  to arbitrary
dimensions;  the deeper  reason  for  that will  become  clear in  the
Appendix.   --  We get  ${\cal P}(r)=1/2$  for  randomly oriented  vectors,
whereas  it  is   becoming  larger  for  the  case   that  ${\mathbf{l}}_2$  is
perpendicular to ${\mathbf{l}}_1$ as well as to $\hat{\mathbf{r}}$. 
\end{enumerate}
Applications of vector marks  can be found in Section~\ref{sec:kerscher_halos}
(Figure~\ref{fig:kerscher_halos-orientation}) where we consider the orientation
of dark matter halos in cosmological simulations. But one can think of
other   applications:  mark   correlation  functions   may   serve  as
orientational  order  parameters  in   liquid  crystals  in  order  to
discriminate between nemetic and  smectic phases (see the contribution
by F.~Schmid and N.~H.~Phuong in  this volume). They can also quantify
the  local orientation  and  order  in liquids  such  as the  recently
measured   five-fold    local   symmetry   found    in   liquid   lead
{}\cite{kerscher_reichert:lead}.  As a further application one could try to measure
the signature of hexatic phases in two-dimensional colloidal
dispersions and in 2D melting scenarios occurring in experiments
and simulations of hard-disk systems (for a review on hard sphere
models see~\cite{kerscher_loewen:lnp}.
%, for instance, the contribution by H. L\"owen to the Lecture
%Notes in Physics, Vol. 554, p. 295-331, Springer-Verlag, Berlin
%2000). 
Finally,  the
orientations   of   anisotropic  channels   in   sandstone  (see   the
contribution  by  C.~Arns et  al.  in  this  volume) are  relevant  for
macroscopic   transport  properties,   therefore   their  quantitative
characterization  in  terms of  mark  correlation  functions might  be
interesting.

\end{enumerate}
Before we move on to applications  a few general remarks are in order:
First,  the definition  of  these mark  characteristics  based on  the
conditional density  ${\cal M}_2(\cdot)$ leads to ambiguities  at $r$ equal
zero as discussed by {}\cite{kerscher_schlather:mark}, but there is no problem
for $r>0$.
-- Furthermore, suitable estimators for  our test quantities are based
on   estimators  for   the  usual   two-point   correlation  function
{}\cite{kerscher_stoyan:fractals,kerscher_capobianco:autocovariance,kerscher_beisbart:luminosity}.
\\
Mark-dependent  clustering  can also  be  defined  at any  $n$-point
level.   Mark-in\-de\-pendent clustering  at  every order  is called  the
random  labelling   property  {}\cite{kerscher_cox:point}.   Mark  correlation
functions based on the $n$-point densities may be used.  For discrete
marks        the        multivariate        $J$-functions        (see
Eq.~\eqref{eq:kerscher_def-multivar-J})   are   an   interesting   alternative,
sensitive  to   higher-order  correlations.   The   random  labelling
property then leads to the relation
\begin{gather}  
J_{i\circ}(r) = J,
\end{gather}
which may be used as a test {}\cite{kerscher_vanlieshout:indices}.

%%%

\section{Describing empirical data: some applications}
\label{sec:kerscher_data}

In many cases already the question whether one or the other type of
dependence as outlined above applies to certain data sets is a
controversial issue.  In the following we will apply our test
quantities to a couple of data sets in order to probe whether there is
an interplay between some objects' marks and their positions in space.
Other applications to biological, ecological, mineralogical,
geological data can be found in
{}\cite{kerscher_stoyan:recent,kerscher_stoyan:fractals,kerscher_ogata:likelihood,kerscher_diggle:statistical}.

\subsection{Segregation effects in the distribution of galaxies}
\label{sec:kerscher_gal}
\begin{figure}
\begin{center}
\epsfig{file=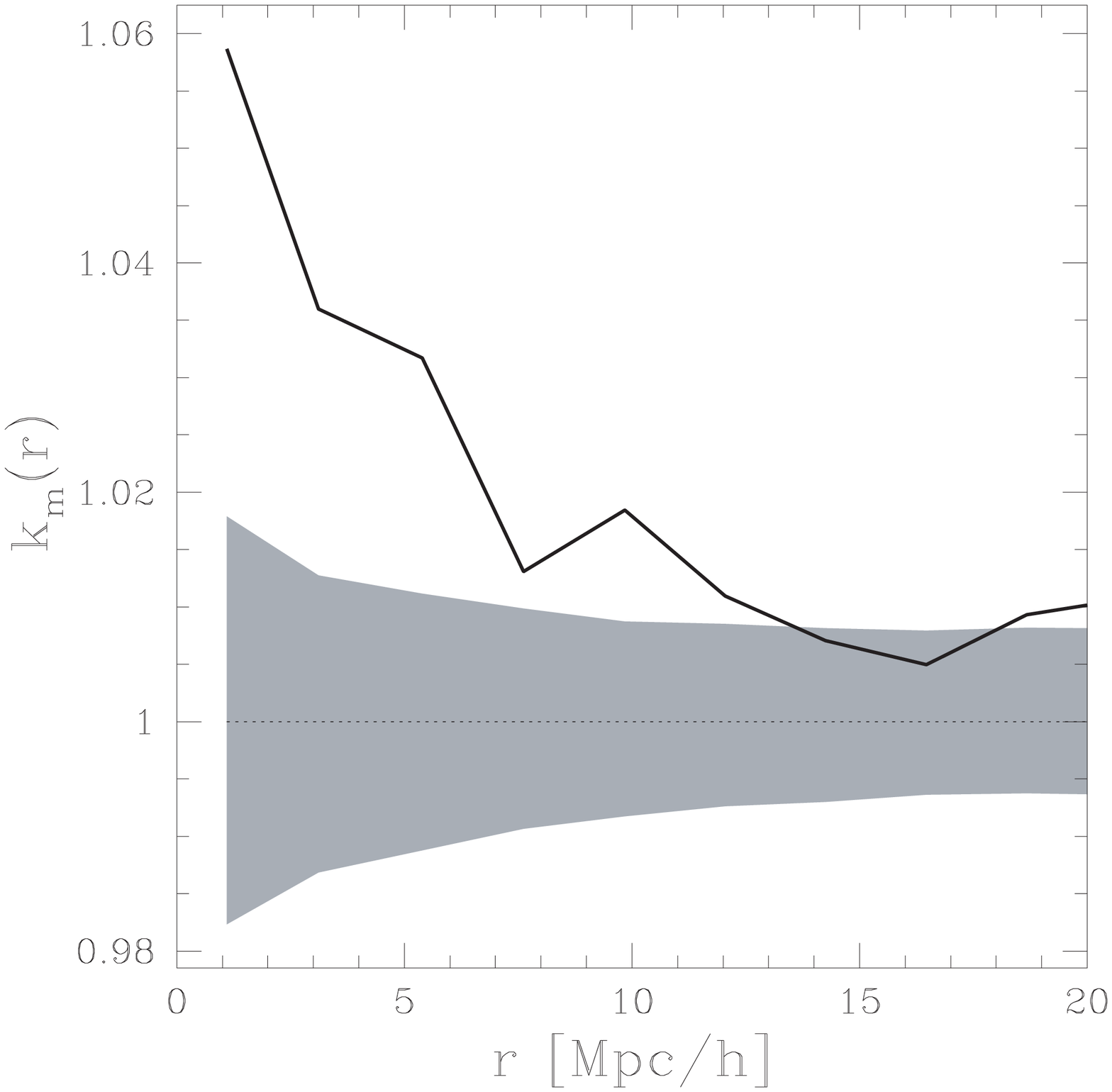,width=5cm}
\epsfig{file=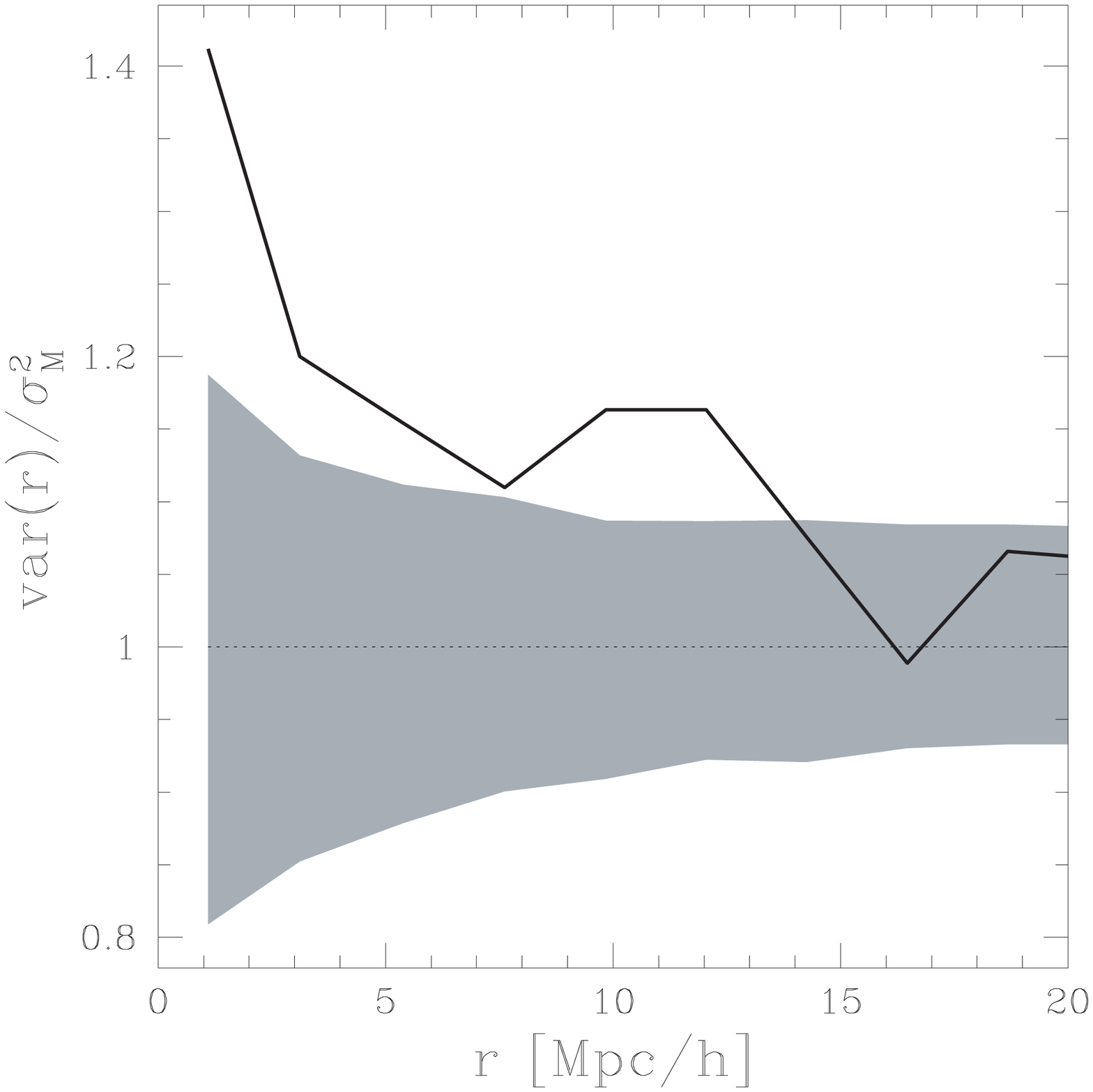,width=4.91cm}
\epsfig{file=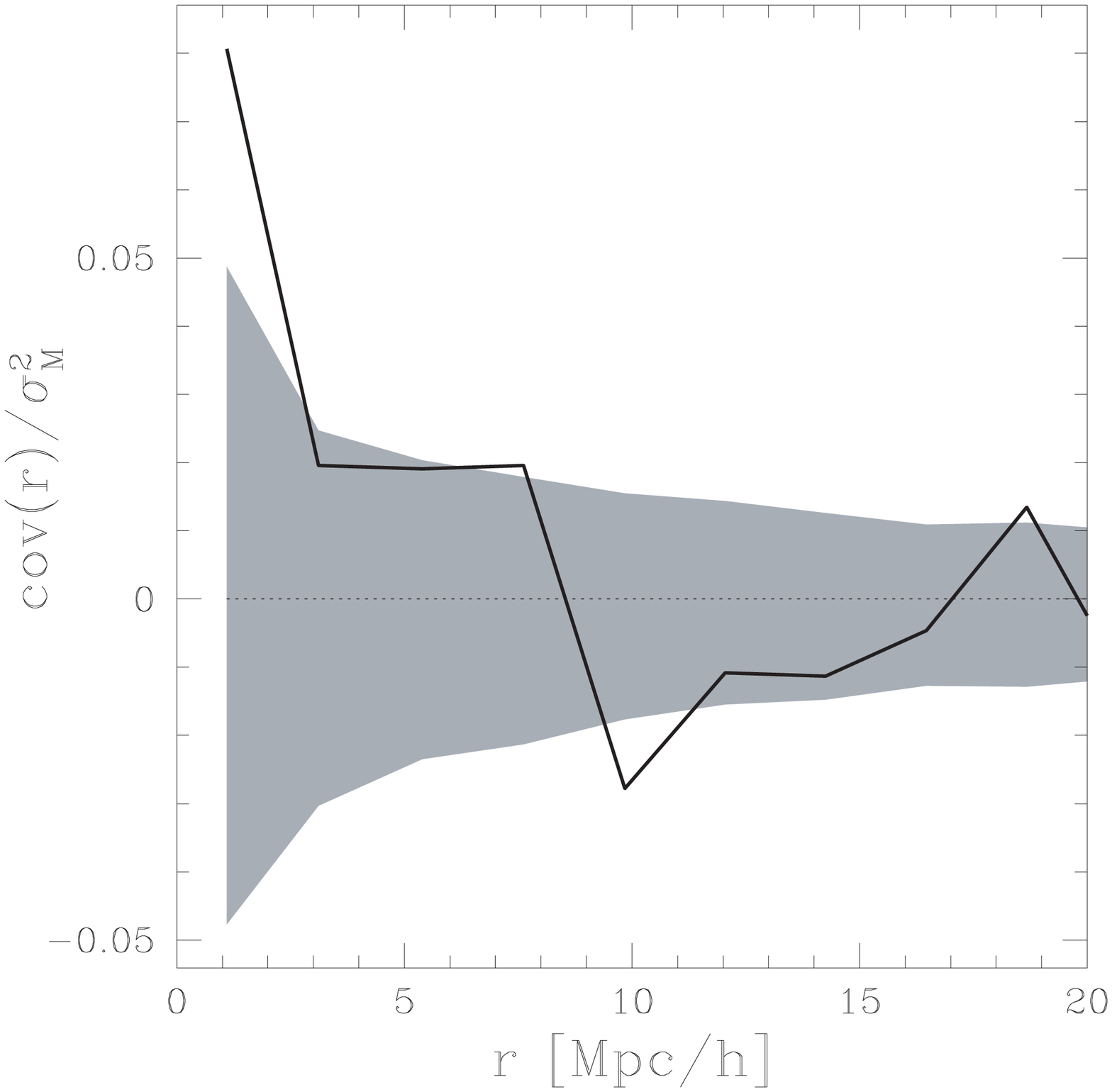,width=5cm}
\end{center}
\caption{The    luminosity-weighted    correlation    functions    for    a
volume-limited subsample  of the SSRS~2  with a depth  of $100{\ifmmode{h^{-1}{\rm Mpc}}\else{$h^{-1}$Mpc}\fi}$.
The shaded  areas denote the  range of one-$\sigma$  fluctuations for
randomized  marks  around  the   case  of  no  mark  segregation.  The
fluctuations         were          estimated         from         1000
reshufflings of the luminosities. \label{fig:kerscher_ssrs-lum} }
\end{figure}

The distribution  of galaxies in  space shows a couple  of interesting
features  and  challenges  theoretical  models  trying  to  understand
cosmological        structure        formation       (see        e.g.\
{}\cite{kerscher_kerscher:statistical}).  There has been a long  debate, whether
and  how  strongly  the   clustering  of  galaxies  depends  on  their
luminosity    and    their     morphological    type    (see,    e.g.\
{}\cite{kerscher_hamilton:evidence,kerscher_hermit:morphology-segregation,kerscher_guzzo:esp}).
The methods which have been used  so far to establish such claims were
based on the spatial two-point correlation function; it was estimated
from different subsamples that were drawn from a catalogue and defined
by morphology  or luminosity.  However, some authors  claimed that the
signal  of luminosity segregation  observed by  others was  a spurious
effect,  caused by  inhomogeneities in  the sample  and  an inadequate
choice       of      the       statistics      {}\cite{kerscher_labini:scale}.
{}\cite{kerscher_beisbart:luminosity}  could show  that methods  based  on the
mark-correlation        functions,       as        discussed       in
Sect.~\ref{sec:kerscher_mark-segregation},     are     not     impaired     by
inhomogeneities, and found a clear signal of luminosity and morphology
segregation.\\
In order to quantify segregation effects in the galaxy distribution we
consider    the    Southern    Sky    Redshift    Survey~2    (SSRS~2,
{}\cite{kerscher_dacosta:southern}),  which maps  a significant  fraction of
the sky and provides us  with the angular sky positions, the distances
(determined via  the redshifts), and some intrinsic  properties of the
galaxies such as their flux and their morphological type.  As marks we
consider either a galaxy's  luminosity estimated from its distance and
flux, or  its morphological  type. In the  latter case  we effectively
divide  our  sample   into  early-type  galaxies  (mainly  elliptical
galaxies)  and  late-type  galaxies  (mainly spirals).  In  order  to
analyze homogeneous  samples, we focus on a  volume-limited sample of
$100{\ifmmode{h^{-1}{\rm Mpc}}\else{$h^{-1}$Mpc}\fi}$  depth\footnote{One ${\ifmmode{{\rm Mpc}}\else{Mpc}\fi}$  equals  roughly $3.26$  million
light  years. The  number  $h$  accounts for  the  uncertainty in  the
measured  Hubble  constant  and  is about  $h\approx 0.65$.   Volume-limited
samples are  defined by  a limiting depth  and a  limiting luminosity.
One considers  only those galaxies  which could have been  observed if
they    were    located    at    the    limiting    depth    of    the
sample.}~\cite{kerscher_beisbart:luminosity}.
\begin{figure}
\begin{center}
\epsfig{file=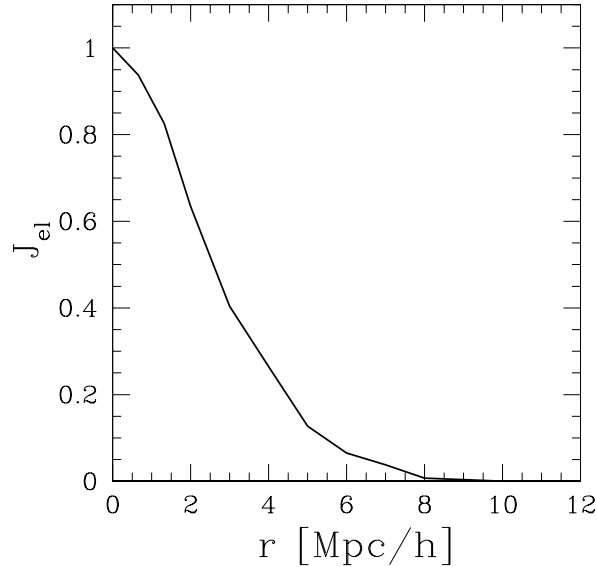,width=8cm}
\end{center}
\caption{\label{fig:kerscher_morph_ind} The $J_{el}$ function of early-type (e)
and late-type (l) galaxies vs.\ the galaxy
separation $r$ in a volume-limited sample of
$100{\ifmmode{h^{-1}{\rm Mpc}}\else{$h^{-1}$Mpc}\fi}$ depth from the SSRS~2 catalogue.}
\end{figure}
\\
In a first step we ask whether the early- and the late-type galaxies
form independent sub-processes.  In Figure~\ref{fig:kerscher_morph_ind} we
show $J_{el}$ as function of the distance $r$ being far away from the
value of one.  Recalling Eq.~\eqref{eq:kerscher_jijone}, we conclude that the
morphological types of galaxies are not distributed
independently on the sky. Not surprisingly, the inequality $J_{el}<1$ indicates
positive interactions between the galaxies of both morphological
types; indeed galaxies attract each other through gravity
irrespective of their morphological types.
\\
After having confirmed the presence of interactions between the
different types of galaxies, we tackle the issue whether the
clustering of galaxies is different for different galaxies.  We
consider the luminosities as marks (see
Fig.~\ref{fig:kerscher_galaxies-circles}).  In Figure~\ref{fig:kerscher_ssrs-lum} we
show some of the mark-weighted conditional correlation functions.  Already at
first glance, they show evidence for luminosity segregation, relevant
on scales up to $15{\ifmmode{h^{-1}{\rm Mpc}}\else{$h^{-1}$Mpc}\fi}$.  To strengthen our claims, we redistribute
the luminosities of the galaxies within our sample randomly, holding
the galaxy positions fixed.  In that way we mimic a marked point
process with the same spatial clustering and the same one-point
distribution of the luminosities, but without luminosity segregation.
Comparing with the fluctuations around this null hypothesis, we see
that the signal within the SSRS~2 is significant.
\\
The  details of the  mark correlation  functions provide  some further
insight into the segregation effects.
The mean mark $k_m(r)>1$ indicates that the luminous galaxies are more
strongly clustered than the dim ones.  Our signal is scale-dependent
and decreasing for higher pair separations.  The stronger clustering
of luminous galaxies is in agreement with earlier claims comparing the
correlation amplitude of several volume-limited samples
{}\cite{kerscher_willmer:southern}.  \\
The $\var(r)$ being larger than the mark variance of the whole sample,
$\sigma^2_M$, shows that on galaxy pairs with separations smaller than
$15{\ifmmode{h^{-1}{\rm Mpc}}\else{$h^{-1}$Mpc}\fi}$ the luminosity fluctuations are enhanced.  The fact that the
mark segregation effect extends to scales of up to $15{\ifmmode{h^{-1}{\rm Mpc}}\else{$h^{-1}$Mpc}\fi}$ is
interesting on its own. In particular, it indicates that galaxy
clusters are not 
the only source of luminosity segregation, since typically galaxy
clusters are of the size of $3{\ifmmode{h^{-1}{\rm Mpc}}\else{$h^{-1}$Mpc}\fi}$.
\\
The signal for the covariance $\cov(r)$, however, could be due to
galaxy pairs inside clusters.  It is relevant mainly on scales up to
$4{\ifmmode{h^{-1}{\rm Mpc}}\else{$h^{-1}$Mpc}\fi}$ indicating that the luminosities on galaxy pairs with small
separations tend to assume similar values. 
-- Our results in part confirm claims by {}\cite{kerscher_benoist:biasing}, who
compared the correlation functions $\xi_2$ for different
volume-limited subsamples and different luminosity classes of the
SSRS~2 catalog (see also {}\cite{kerscher_benoist:biasinghigher}). 

%%%
\subsection{Orientations of dark matter halos}
\label{sec:kerscher_halos}
\begin{figure}
\begin{center}
\epsfig{file=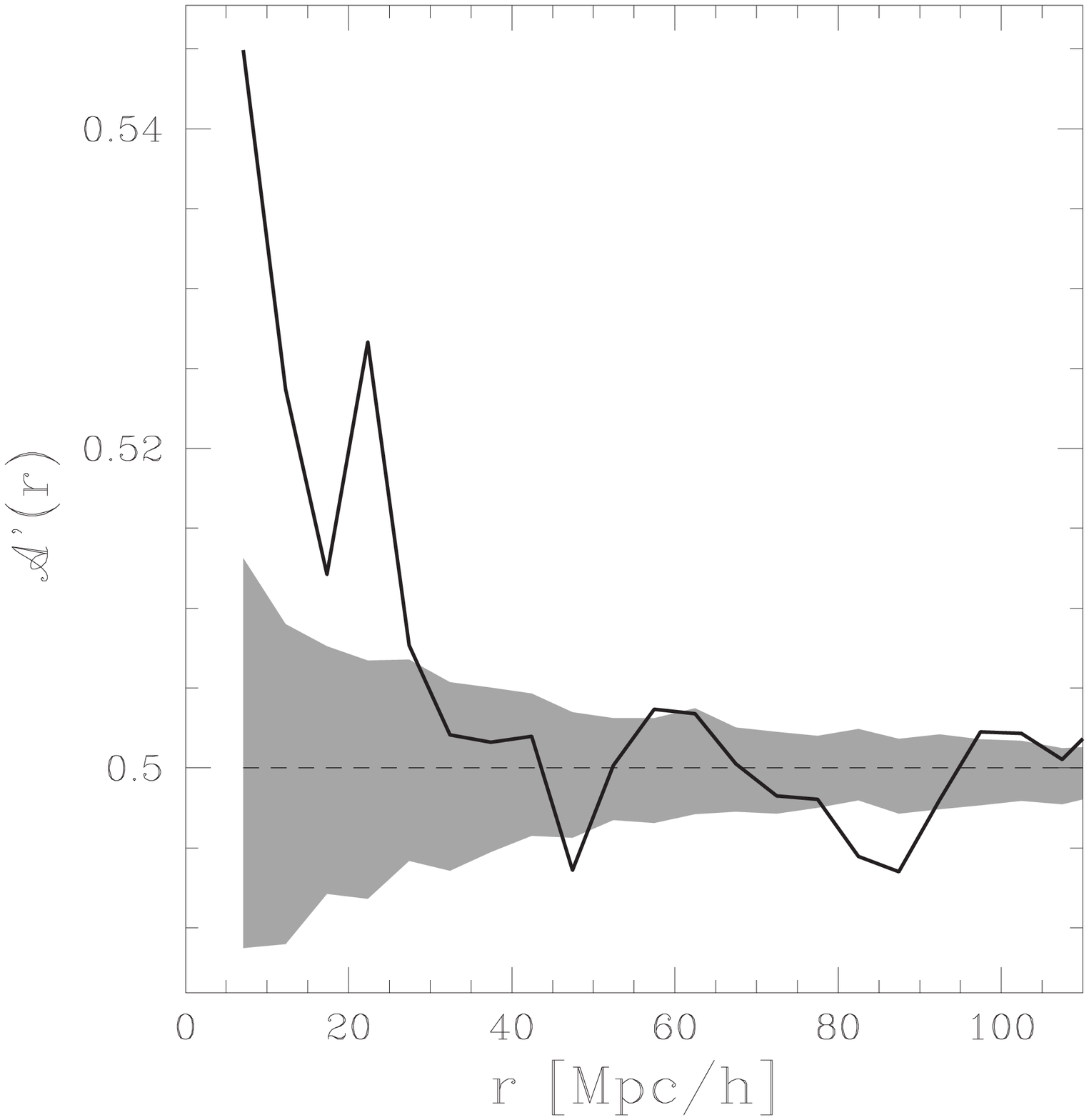,width=4.9cm}
\epsfig{file=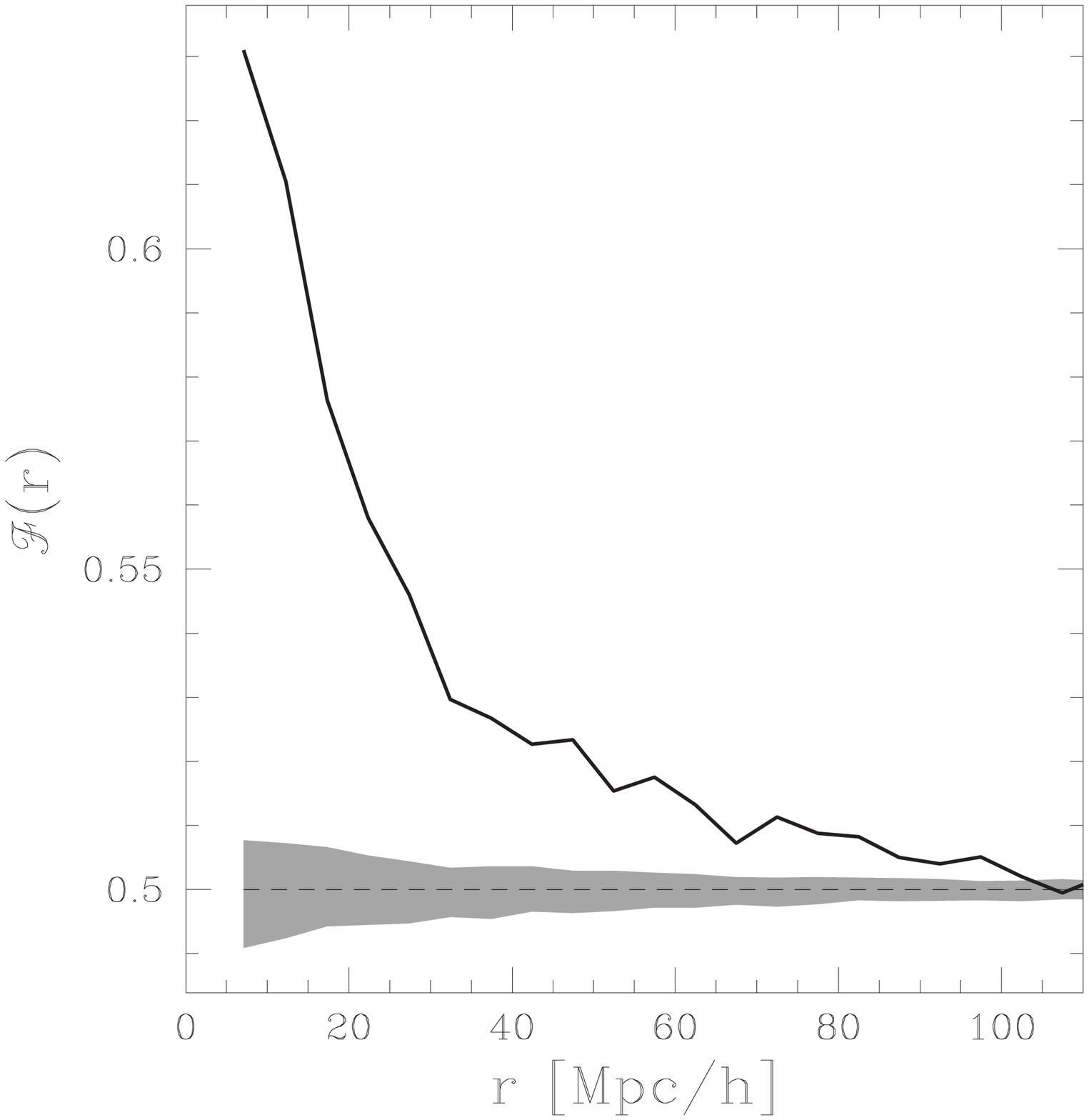,width=4.9cm}
\epsfig{file=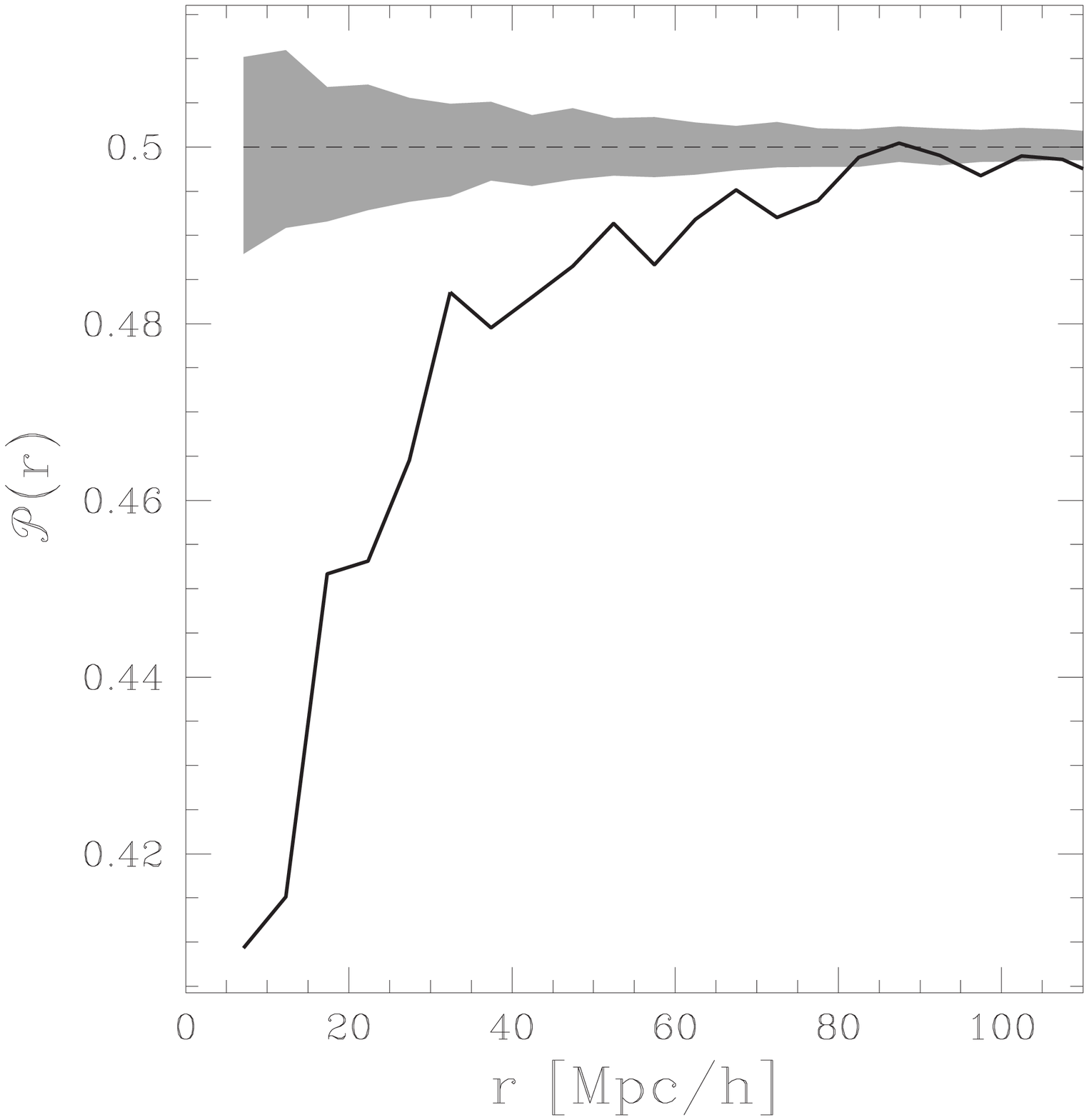,width=4.9cm}
\end{center}
\caption{The correlations of halo orientations in numerical simulations. The orientation of each dark matter halo, specified by the
direction of the major axis ${\mathbf{l}}$ of the mass ellipsoid, is used as a
vector mark. The dashed area is obtained by randomizing the
orientations among the halos.
\label{fig:kerscher_halos-orientation}}
\end{figure}

Many  structures found  in the  Universe such  as galaxies  and galaxy
clusters  show   anisotropic  features.   Therefore   one  can  assign
orientations to them and ask whether these orientations are correlated
and form coherent patterns.  Here we discuss a similar question on the
base  of  numerical  simulations   of  large  scale  structure  (e.g.,
{}\cite{kerscher_bertschinger:simulations,kerscher_klypin:numericalI}).   \\  In such
simulations  the  trajectories of  massive  particles are  numerically
integrated. These  particles represent the dominant  mass component in
the Universe, the dark matter.  Through gravitational instability high
density peaks  (``halos'') form in the distribution  of the particles;
these halos are likely to be the places where galaxies originate.
In the following we will report on
alignment correlations between such halos
{}\cite{kerscher_faltenbacher:halos}, for a further application of mark
correlation functions in this field see {}\cite{kerscher_gottloeber:merger}.
\\
The halos used  by {}\cite{kerscher_faltenbacher:halos} stem from a
$N$-body  simulation  in  a  periodic  box  with  a  side  length  of
500{\ifmmode{h^{-1}{\rm Mpc}}\else{$h^{-1}$Mpc}\fi}. The initial and  boundary conditions were fixed according to
a $\Lambda$CDM cosmology (for a discussion of cosmological models
see \cite{kerscher_peebles:principles,kerscher_coles:cosmology}). Halos were
identified using a friend-of-friends algorithm in the dark matter
distribution. Not all of the halos found were taken into account; rather the mass range and
the spatial  number density  of the selected halos  were chosen to  resemble the
properties of observed galaxy  clusters in the \textsc{Reflex} catalogue
{}\cite{kerscher_boehringer:reflexI}.   Typically our halos show  a  prolate
distribution of their dark matter particles.
\\
For each halo the direction of the elongation is determined from the
major axis of the mass-ellipsoid. This leads to a marked point set
where the orientation ${\mathbf{l}}_i$ is attached to each halo position ${\mathbf{x}}_i$ as
a vector mark with $|{\mathbf{l}}_i|=1$. Details can be founds in
{}\cite{kerscher_faltenbacher:halos}.
\\
In Fig.~\ref{fig:kerscher_halos-orientation} the vector-mark correlation functions  as defined in Eqs.~\eqref{eq:kerscher_align}, \eqref{eq:kerscher_filament},
and  \eqref{eq:kerscher_planar} are shown.   Since only  the orientation  of the
mass    ellipsoids    can    be    determined,   we    use    ${\cal A}'(r)$
(Eq.~\ref{eq:kerscher_def-vector-corr-absolute}) instead of ${\cal A}(r)$.
The signal in ${\cal A}'(r)$ indicates  that pairs of halos with a distance
smaller than 30{\ifmmode{h^{-1}{\rm Mpc}}\else{$h^{-1}$Mpc}\fi}\  show a tendency of parallel  alignment of their
orientations ${\mathbf{l}}_1,{\mathbf{l}}_2$. The deviation from a pure random alignment
is in the percent range but clearly outside the random fluctuations.
The alignment of the halos' orientations ${\mathbf{l}}_1,{\mathbf{l}}_2$ with the
connecting vector $\hat{\mathbf{r}}$ quantified by ${\cal F}(r)$ is significantly
stronger; it is particularly interesting that this alignment effect
extends to scales of about  100{\ifmmode{h^{-1}{\rm Mpc}}\else{$h^{-1}$Mpc}\fi}.
\\
In a qualitative picture this may be explained by halos aligned
along the filaments of the large scale structure. Indeed such
filaments are prominent features found in the galaxy distribution
{}\cite{kerscher_huchra:cfa2s1} and in $N$-body simulations
{}\cite{kerscher_melott:generation}, often with a length of up to ~100{\ifmmode{h^{-1}{\rm Mpc}}\else{$h^{-1}$Mpc}\fi}.
The lowered ${\cal P}(r)$ indicates that the volume of the rhomboid given
by ${\mathbf{l}}_1,{\mathbf{l}}_2$ and $\hat{\mathbf{r}}$ is reduced for halo pairs with a
separation below 80{\ifmmode{h^{-1}{\rm Mpc}}\else{$h^{-1}$Mpc}\fi}. Already a preferred alignment of
${\mathbf{l}}_1,{\mathbf{l}}_2$ along $\hat{\mathbf{r}}$ leads to such a reduction, similar to a
plane-like arrangement of ${\mathbf{l}}_1,{\mathbf{l}}_2,\hat{\mathbf{r}}$. For the halo
distribution the signal in ${\cal P}(r)$ seems to be dominated by the
filamentary alignment.
\\ The  question whether  there are non-trivial  orientation patterns
for galaxies  or galaxy clusters has  been discussed for  a long time.
{}\cite{kerscher_binggeli:shape}  reported  a  significant  alignment  of  the
observed       galaxy       clusters       out      to       100{\ifmmode{h^{-1}{\rm Mpc}}\else{$h^{-1}$Mpc}\fi}.
{}\cite{kerscher_struble:new,kerscher_struble:new-erratum},  however claimed  that this
effect   is  small   and   likely  to   be   caused  by   systematics;
{}\cite{kerscher_ulmer:major} find no indication for alignment effects at all.
Subsequently  several  authors  purported   to  have  found  signs  of
alignments in  the galaxy and  galaxy cluster distribution  (see e.g.\
{}\cite{kerscher_djorgovski:coherent,kerscher_lambas:statistics,kerscher_fuller:alignments,kerscher_heavens:intrinsic}).
Our  Fig.~\ref{fig:kerscher_halos-orientation}   shows  that  from  simulations
 significant  large-scale  correlations are  to  be  expected in  the
 orientations  of galaxy clusters,  in agreement  with the  results by
 {}\cite{kerscher_binggeli:shape}.  These  results  are  also supported  by  a
 simulation study carried out by {}\cite{kerscher_onuora:alignment}.

\subsection{Martian Craters}
\label{sec:kerscher_martian}
\begin{figure}
\begin{center}
\epsfig{file=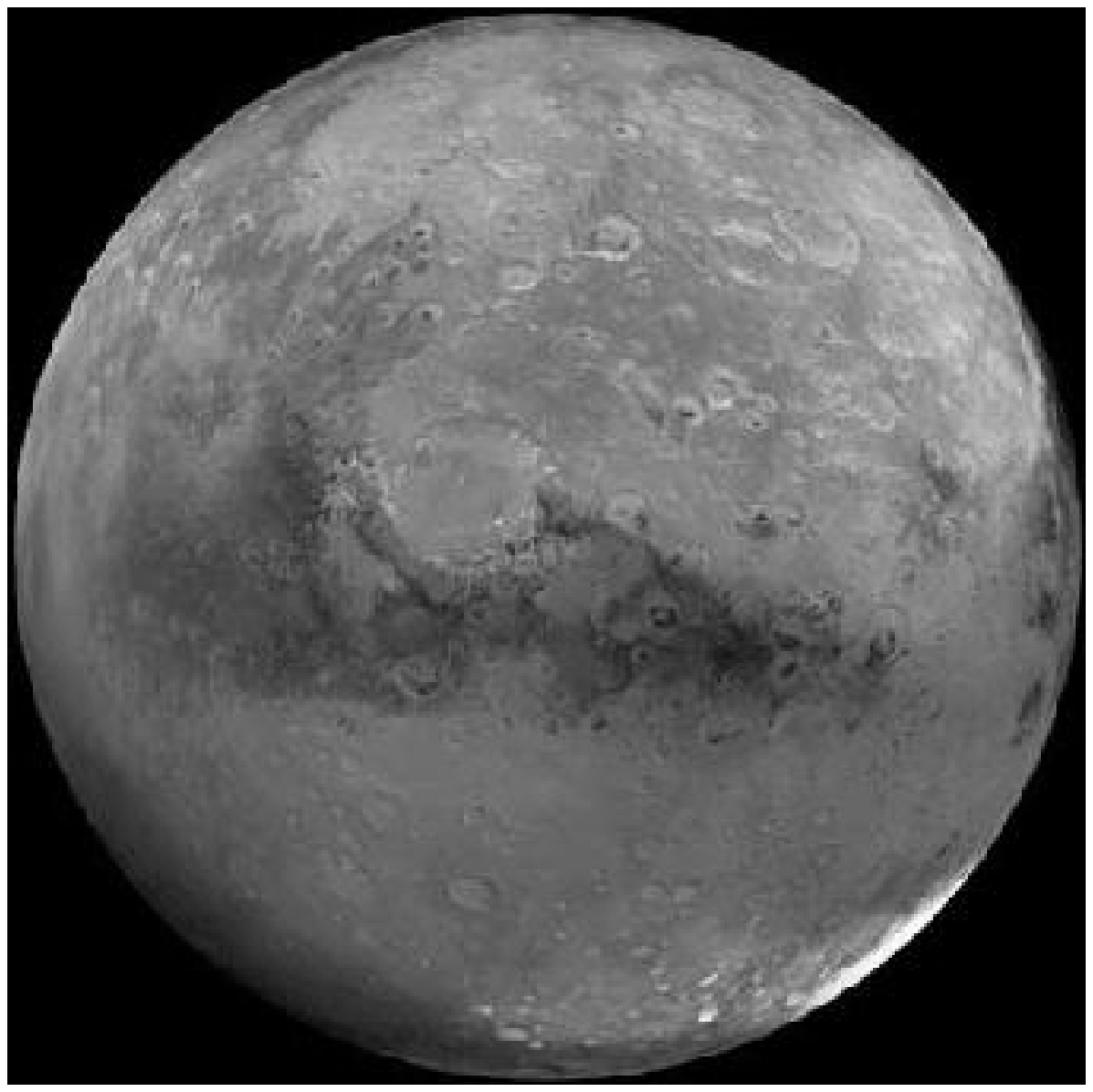,width=6cm}
\epsfig{file=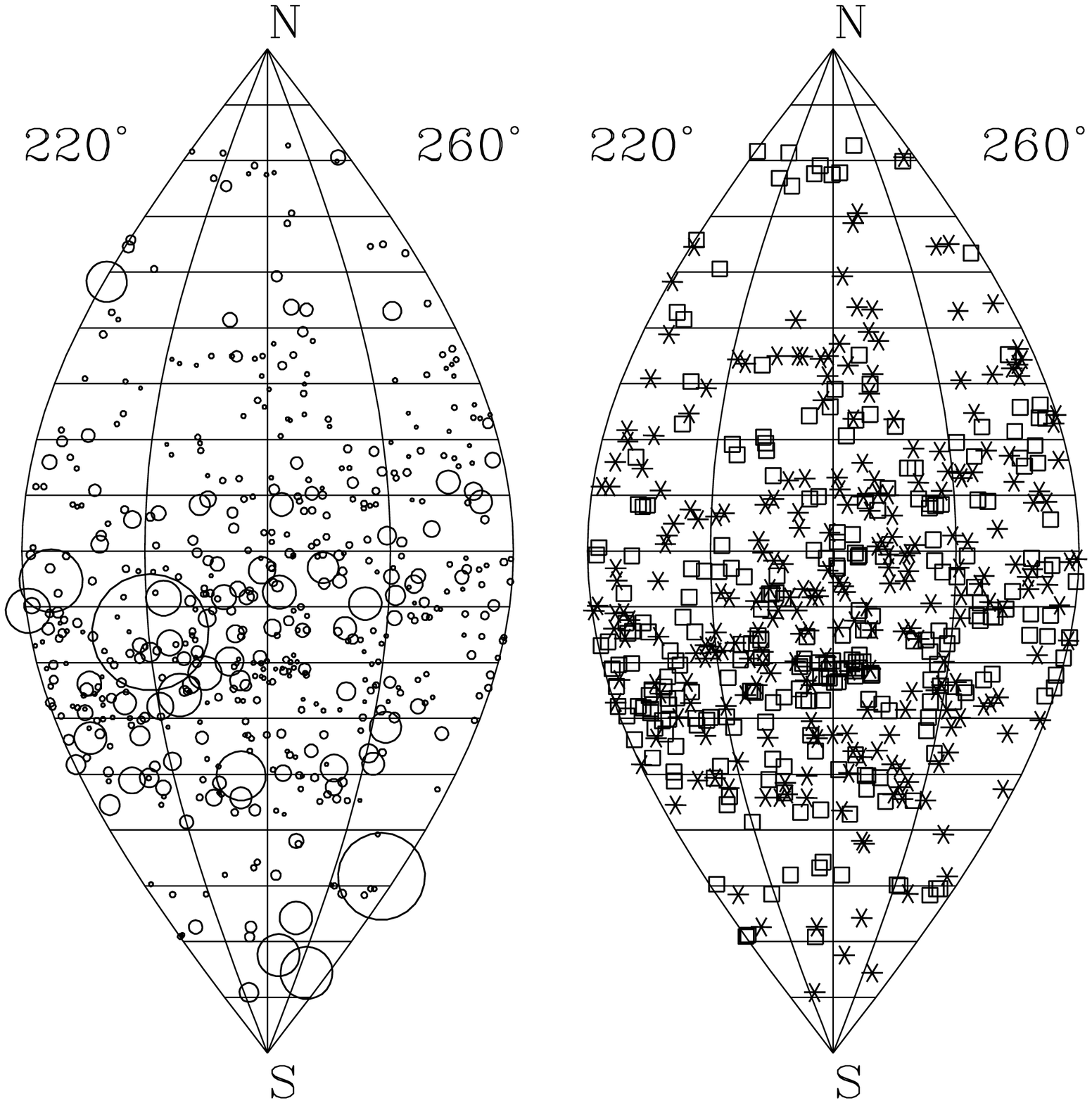,width=6cm}
\end{center}
\caption{ The Martian surface with its craters. Whereas the left panel
(from         http://pds.jpl.nasa.gov/planets/captions/mars/schiap.htm)
illustrates  the  various  geological  settings  to be  found  on  the
planet's surface, the other panels focus on a small patch and show the
craters  together with  their radii  (middle  panel, the  size of  the
symbols are  proportional to  the radii of  the craters)  and together
with the craters' types (right panel, simple morphology as quadrangles
and more complex craters as  stars). The latter viewgraphs rely on the
data by {}\cite{kerscher_barlow:martian}.
\label{fig:kerscher_mars_eg}}
\end{figure}

\begin{figure}
\begin{center}
\epsfig{file=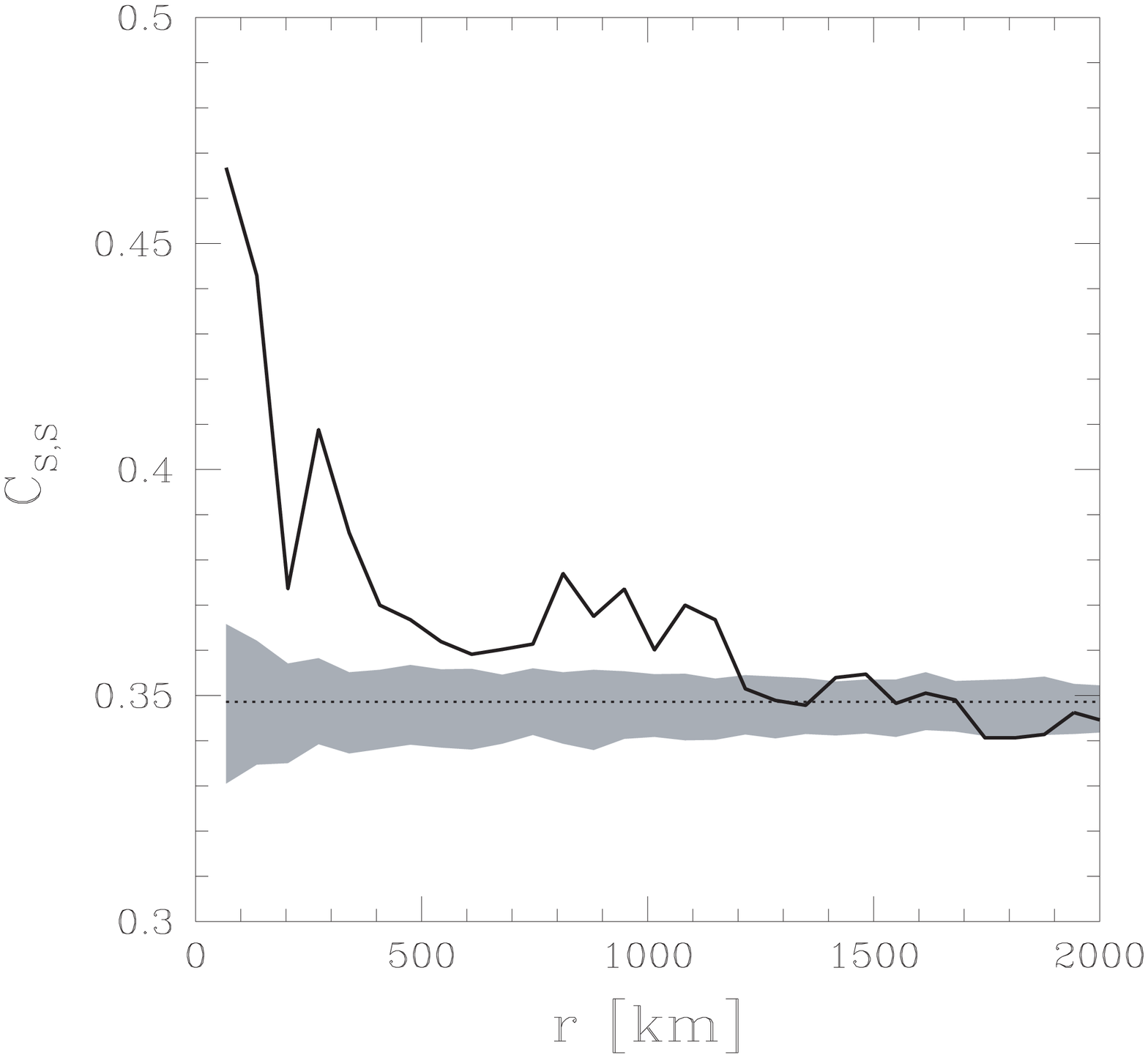,width=5cm}
\epsfig{file=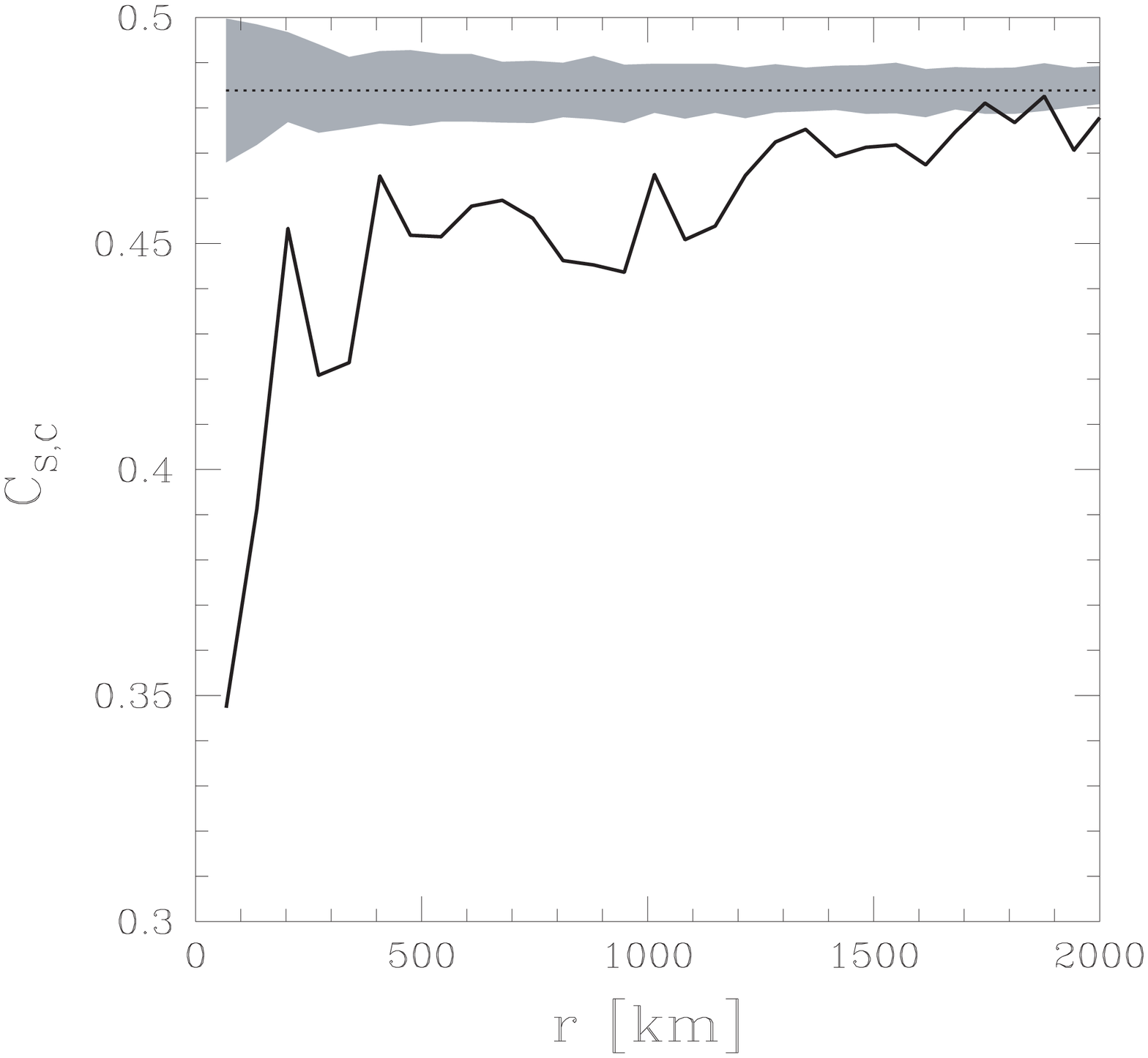,width=5cm}
\epsfig{file=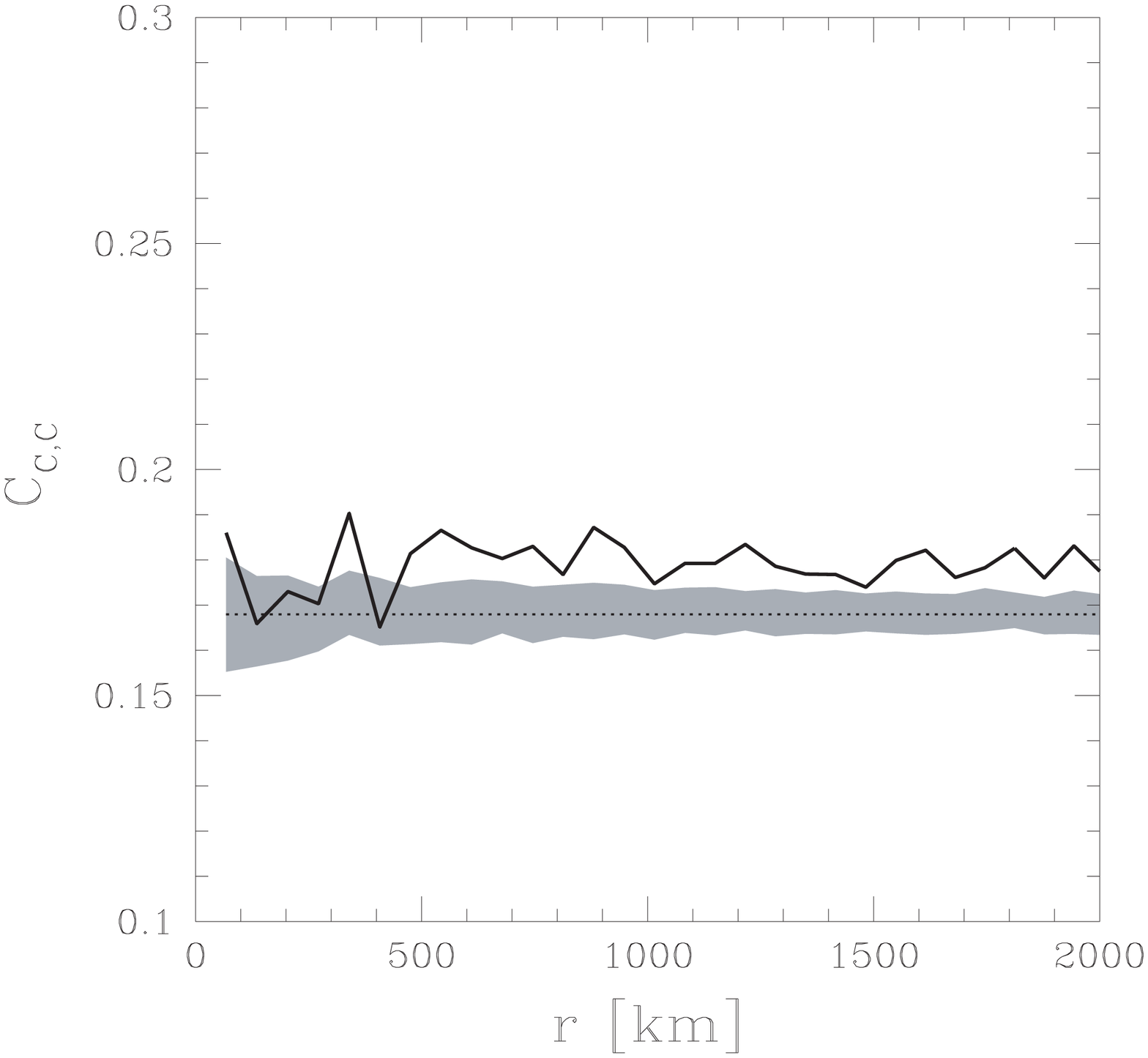,width=5cm}
\end{center}
\caption{\label{fig:kerscher_mars-crosscorr} The conditional cross-correlation
functions for Martian craters. We  split the sample of craters into two
broad classes according to their ejecta types: simple morphologies (S)
consisting of SL  and DL types, and complex  morphologies (C) with all
other types (see  {}\cite{kerscher_barlow:martian} for details). The results
indicate, that at  scales up to about $1500$ km  the clustering of the
simple  craters is  enhanced  at expense  of  cross correlations.  The
shaded  areas  denote the  one-$\sigma$  fluctuations for  randomized
marks estimated from 100 realizations of the mark reshuffling. }
\end{figure}

Let  us now turn  to another,  still astrophysical,  but significantly
closer object: the Mars (see Figure~\ref{fig:kerscher_mars_eg}).  Many planets'
surfaces display impact craters with diameters up to $\sim260\,$km and
a broad range  of inner morphologies. These craters  are surrounded by
ejecta  forming different types  of patterns.   The craters  and their
ejecta  are likely  to  be  caused by  asteroids  and periodic  comets
crossing the planets' orbits,  falling down onto the planet's surface,
and  spreading some  of the  underlying surfaces  material  around the
original impact  crater.  A  variety of different  crater morphologies
and a  wide range of ejecta  patterns can be found. In principle,  either the
different impact
objects (especially their energies)  or the various surface  types of the planet  may explain  the repertory of patterns observed.
Whereas the energy variations of impact objects do not cause any
peculiarities in the spatial distribution of the craters (apart from a
possible latitude dependence), geographic inhomogeneities are expected
to originate inhomogeneities in the craters' morphological properties.
\\
We try to answer the question for the ejecta patterns' origin using
data collected by {}\cite{kerscher_barlow:martian} who already found
correlations between crater characteristics and the local surface type
employing geologic maps of the Mars. Complementary to their approach,
we investigate two-point properties without any reference to geologic
Mars maps.  We restrict ourselves only to craters which have a
diameter larger than $8$ km and whose ejecta pattern could be
classified, ending up with $3527$ craters spread out all over the
Martian surface. We use spherical distances for our analysis of pairs.
\\
In a first step we divide the ejecta patterns into two broad classes
consisting of either the simple patterns (single and double lobe
morphology, i.e.  SL and DL in terms of the classification by
{}\cite{kerscher_barlow:martian}; we speak of ``simple craters'') or the
remaining, more complex configurations (``complex craters'').  Using
our conditional cross correlation functions $C_{ij}$ as defined in
Equation~\eqref{eq:kerscher_cond-crosscorr}, we see a highly significant signal
for mark correlations (Figure~\ref{fig:kerscher_mars-crosscorr}).  At small
separations, crater pairs are disproportionally built up of simple
craters at the expense of cross correlations. This can be explained
assuming that crater formation depends on the local surface type: if
the simple craters are more frequent in certain geological
environments than in others, then there are also more pairs of them to
be found as far as one focuses on distances smaller than the typical
scale of one geological surface type.  Cross pairs are suppressed,
since typical pairs with small separations belong to one geological
setting where the simple craters either dominate or do not. Only a
small, positive segregation signal occurs for the complex craters.
Hence our analysis indicates that the broad class of complex craters
is distributed quite homogeneously over all of the geologies. On top
of this there are probably simple craters, their frequency significantly
depending on the surface type.
\\
If the ejecta patterns were independent of the surface, no mark
segregation could be observed (other sources of mark segregation are
unlikely, since the Martian craters are a result of a long bombardment
history diluting any eventual peculiar crater correlations). In this
sense, the  signal observed indicates a surface-dependence of
crater formation. This result is remarkable, given that we did not use
any geological information on the Mars at all. The picture emerging
could be described using the random field model, where a field
(here the surface type) determines the mark of the points (see below).
\begin{figure}
\begin{center}
\epsfig{file=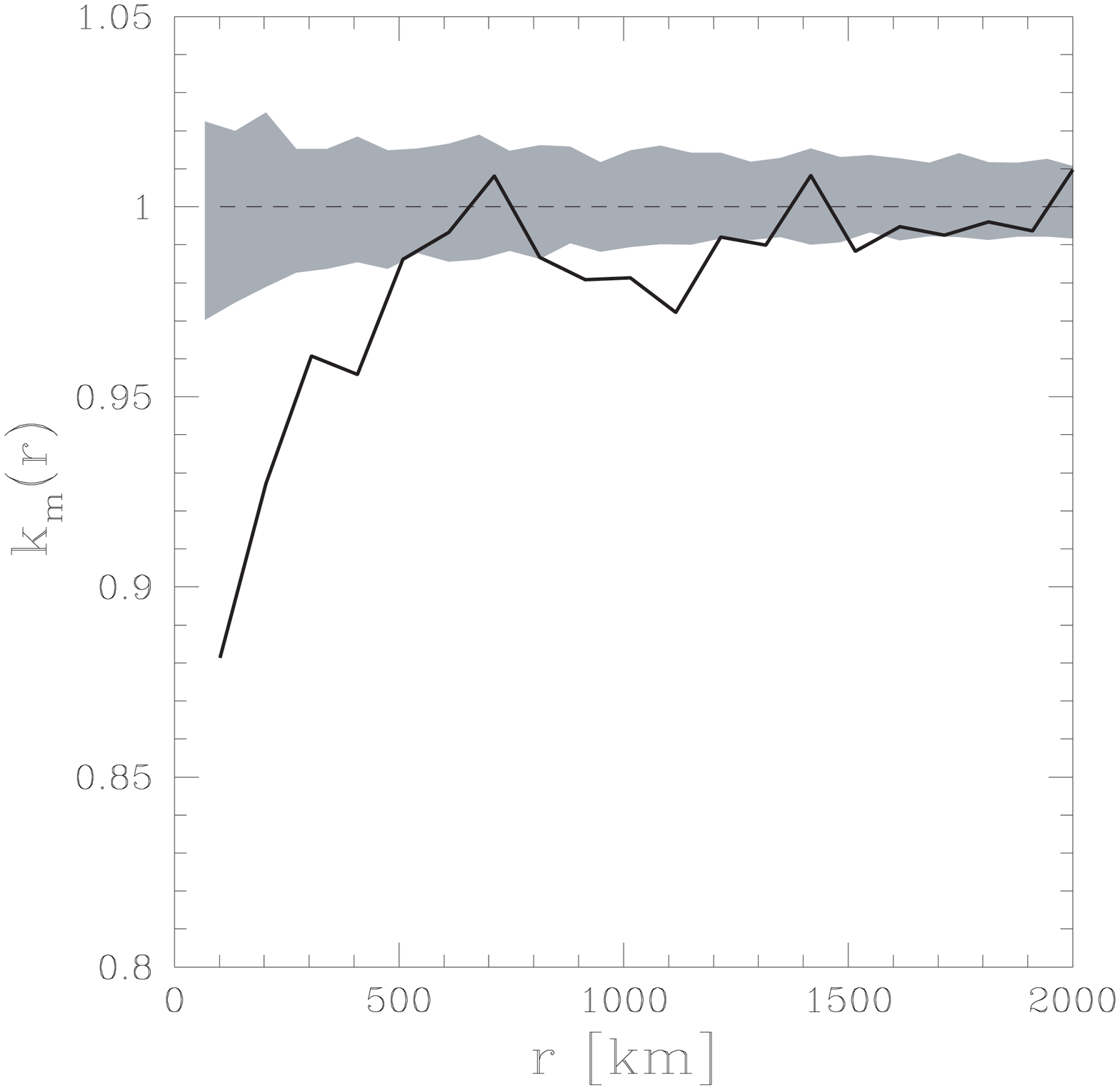,width=5cm}
\epsfig{file=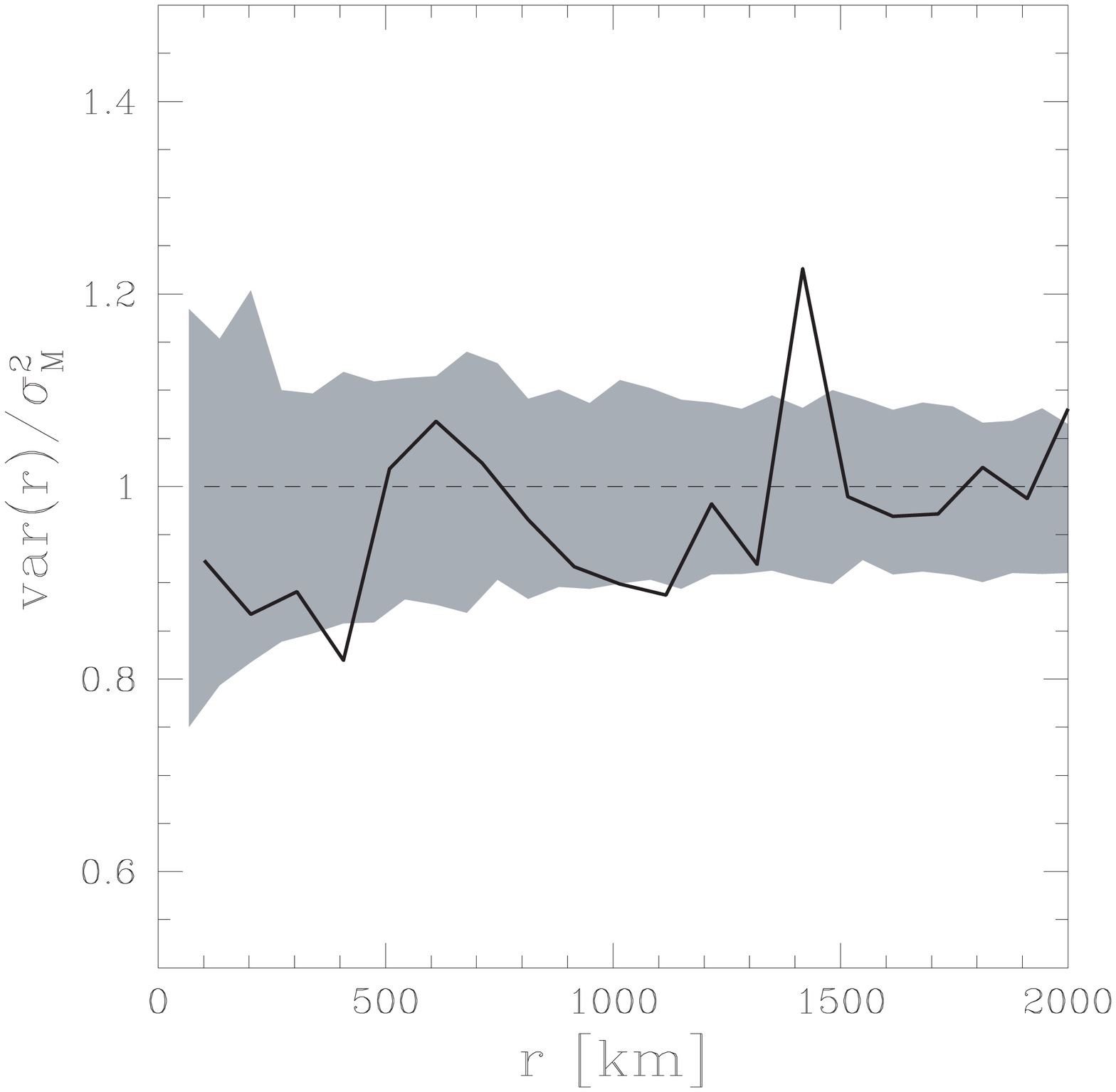,width=4.93cm}
\epsfig{file=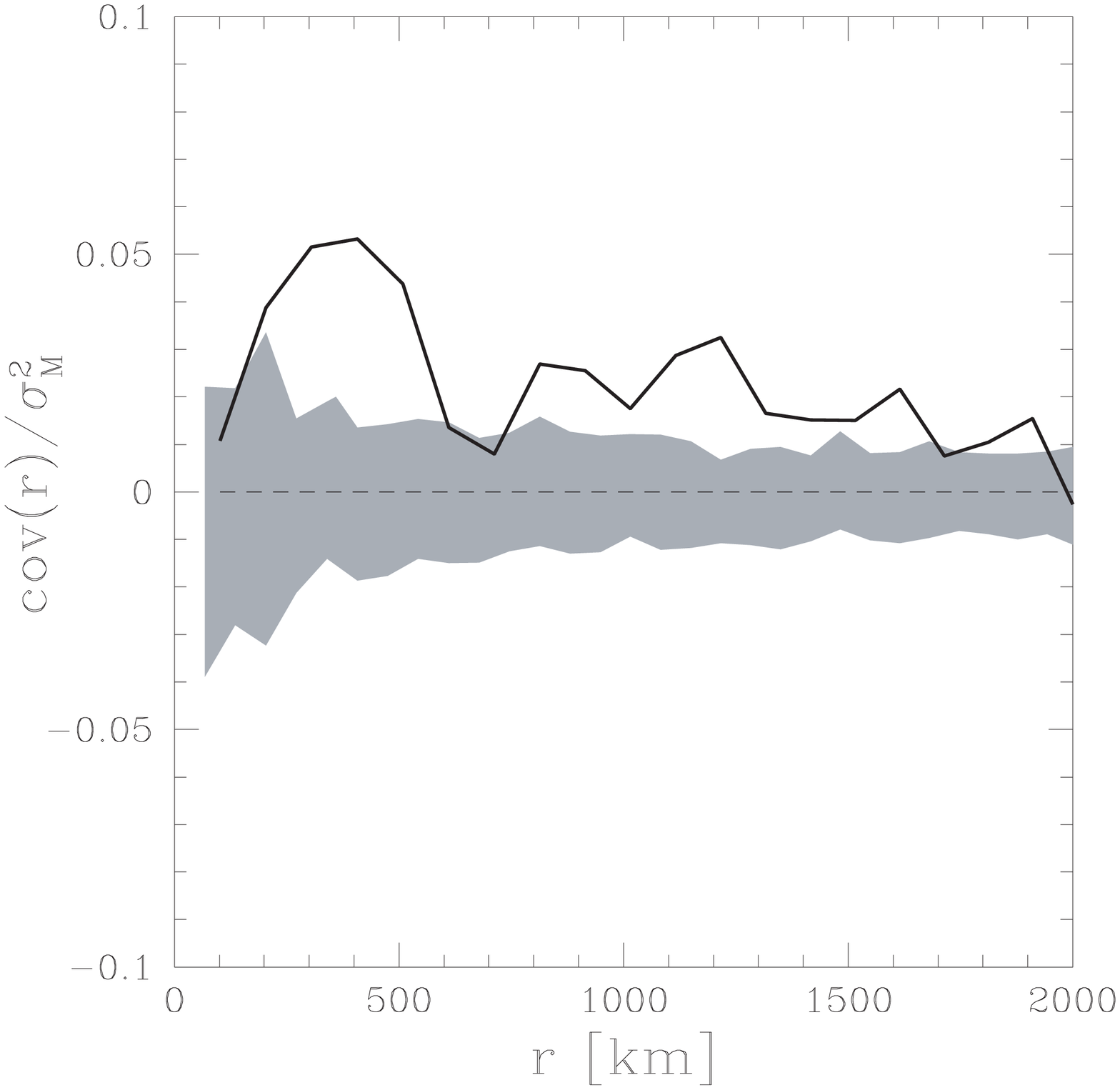,width=5cm}
\end{center}
\caption{The  radius-weighted correlation  functions  for craters  on
Mars. The radius  of each crater serves as a  mark. $r$ is
the  spherical distance.    The  shaded   areas  denote   the  one-$\sigma$
fluctuations    for    randomized    marks    estimated    from    100
realizations of the mark reshuffling. \label{fig:kerscher_mars-markcorr} }
\end{figure}
\\
In  a second  step,  we  analyze the  interplay  between the  craters'
diameters and their spatial clustering.   Now the diameter serves as a
continuous mark. The  results in Figure~\ref{fig:kerscher_mars-markcorr} show a
clear signal for mark segregation in $k_m$ and $\cov$ at small scales.
The latter signals that pairs with  separations in a broad range up to
$1700$ km  tend to have similar  diameters; this is  in agreement with
the  earlier picture: as  {}\cite{kerscher_barlow:martian} showed,  the simple
craters   are  mostly   small-sized.  Pairs   with   relatively  small
separations  thus often  stem  from the  same  geological setting  and
therefore have similar diameters and similar morphological type.
\\
Also the signal of $k_m$ seems to support this picture: since the
simple craters are more strongly clustered than the other ones and
since they have smaller diameters, one could expect $k_m<1$.  As we
shall see in Sect.~\ref{sec:kerscher_models}, however, a $k_m\ne1$ contradicts the
random field model; therefore, the mark-dependence on the underlying
surface type (thought of as a random field) cannot account for the
signal observed. Thus, we have to look for an alternative explanation:
it seems reasonable, that, whenever a crater is found somewhere, no
other crater can be observed close nearby (because an impact close to
an existing crater will either destroy the old one or cover it with
ejecta such that it is not likely to be observed as a crater).  This
results in a sort of effective hard-core repulsion.  This repulsion
should be larger for larger craters.  Thus, pairs with very small
separations can only be formed by small craters, therefore $k_m<1$ for
tiny $r$.  The scale beyond which $k_m(r)\sim 1$ should somehow be
hidden within the crater diameter distribution.  Indeed, at about
$500$ km the segregation vanishes, which is about twice the largest
diameter in our sample.  Taking into account that the ejecta patterns
extend beyond the crater, this seems to be a reasonable agreement. As
shown in Sect.~\ref{sec:kerscher_boolean-depletion} a model based on these
consideration is able to produce such a depletion in the $k_m(r)$.
This effect could also in turn explain part of the cross
correlations observed earlier in Figure~\ref{fig:kerscher_mars-crosscorr}.  A
similar effect is to be expected for the mark variance. Close pairs
are only accessible to craters with a smaller range of diameters;
therefore, their variance is diminished in comparison to the whole
sample. However, an effect like this is barely visible in the data.
\\
Altogether, the crater distribution is dominated by two effects: the
type of the ejecta pattern and the crater diameter depend on the
surface, in addition, there is a sort of repulsion effect on small
scales.

\subsection{Pores in Sandstone}
\label{sec:kerscher_sandstone}

\begin{figure}
\begin{center}
\epsfig{file=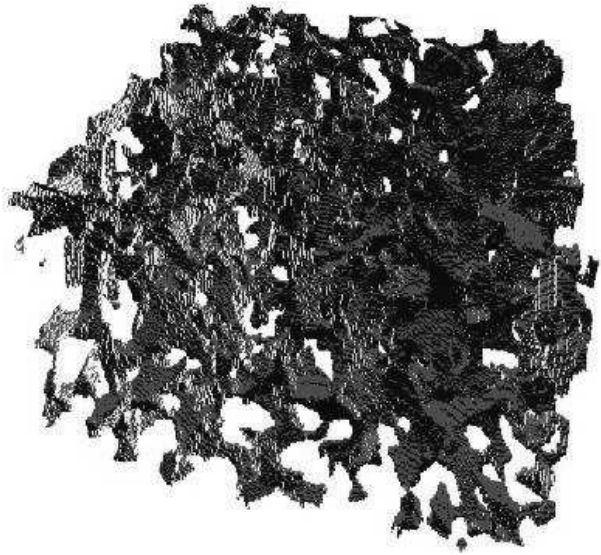,width=8cm}
\end{center}
\caption{The pores within a Fontainbleau sandstone sample.  Note, that
this is a negative image, where  the pores are displayed in grey. The
geometrical features of the pore network are important for macroscopic
properties of the stone. In this sample the pores occupy $13\%$ of the
volume. The size of the whole sample shown is about $1.5$
mm$^3$ (Courtesy M.~Knackstedt).
\label{fig:kerscher_sandstone_eg}}
\end{figure}
\begin{figure}
\begin{center}
\epsfig{file=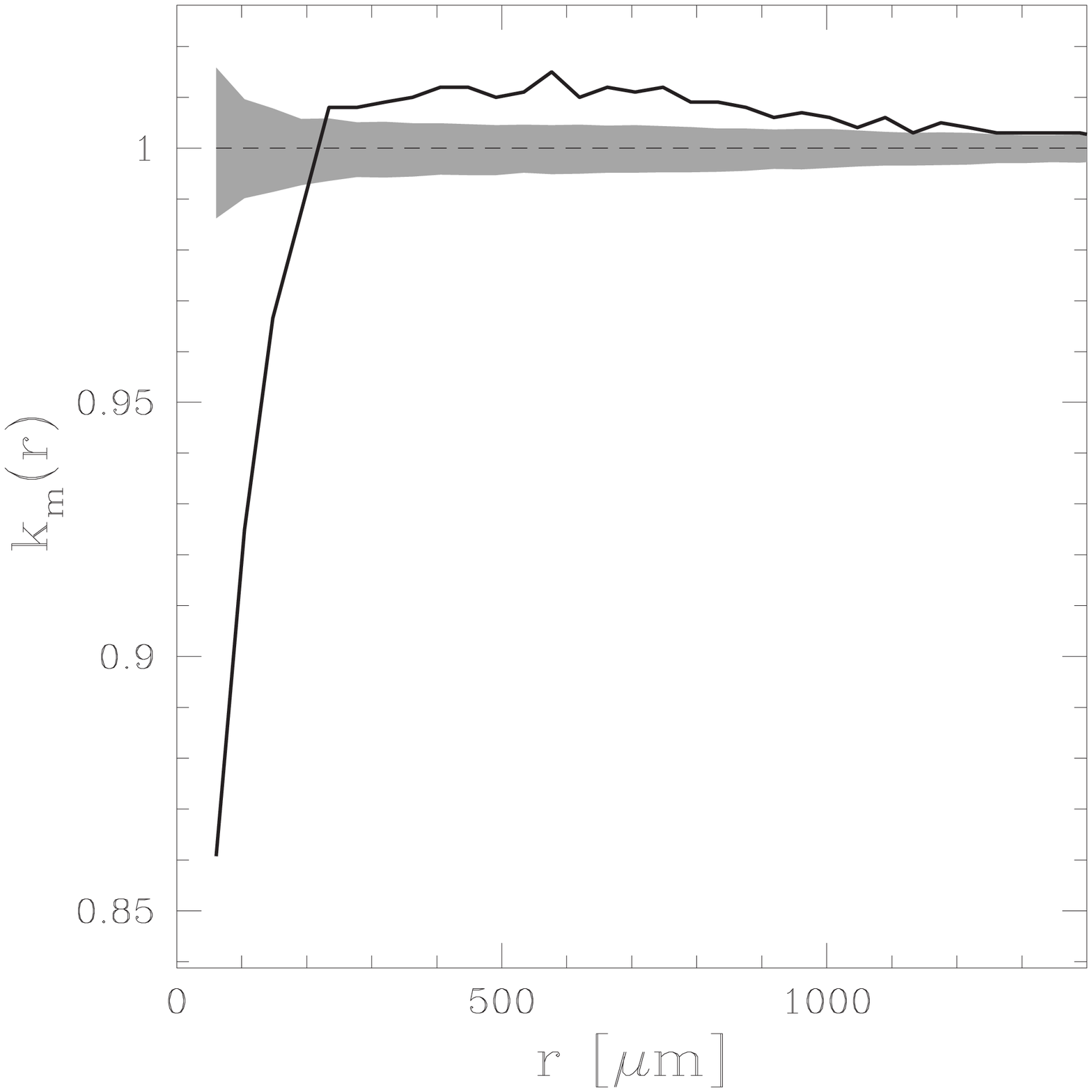,width=5cm}
\epsfig{file=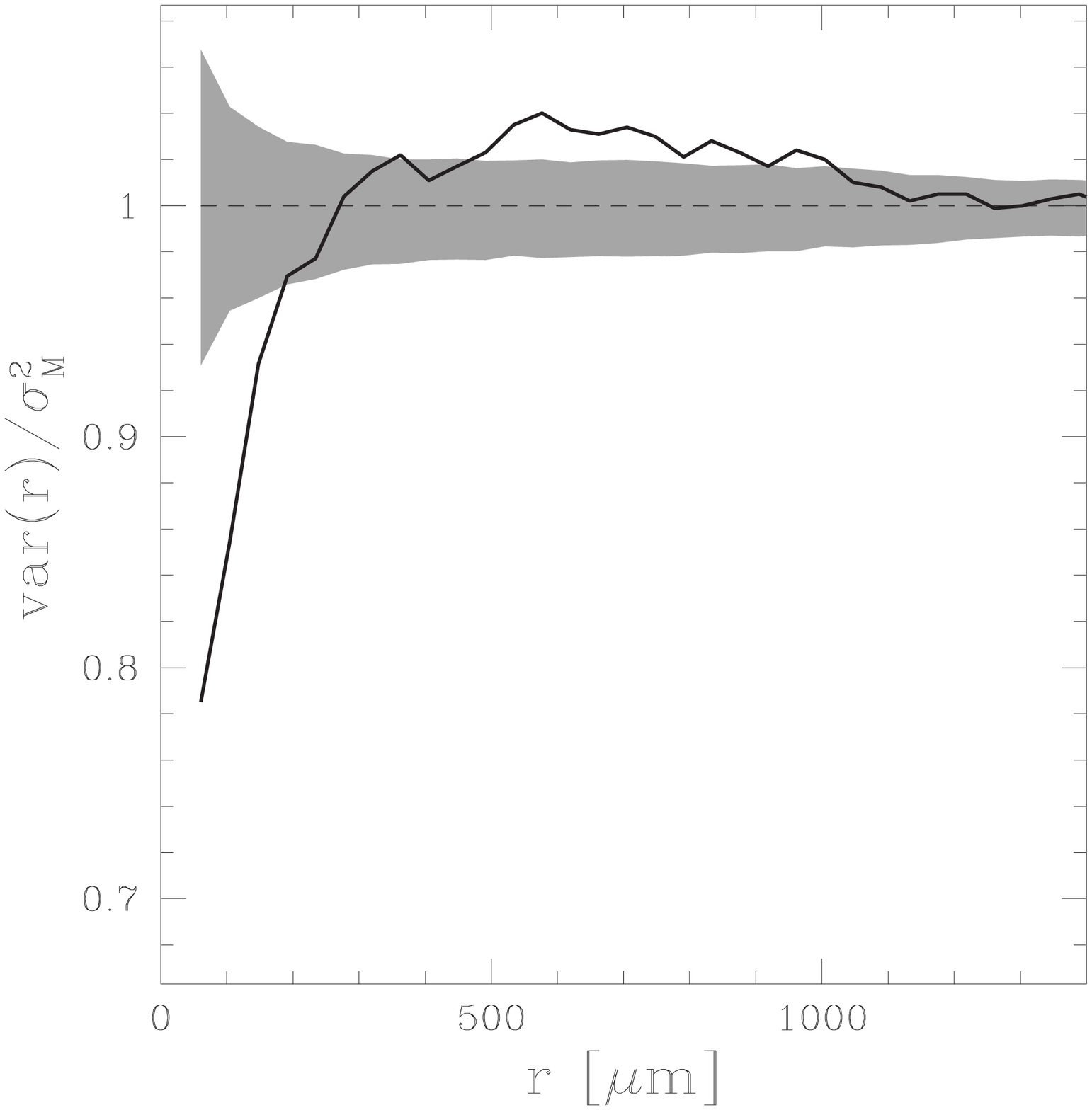,width=5cm}
\epsfig{file=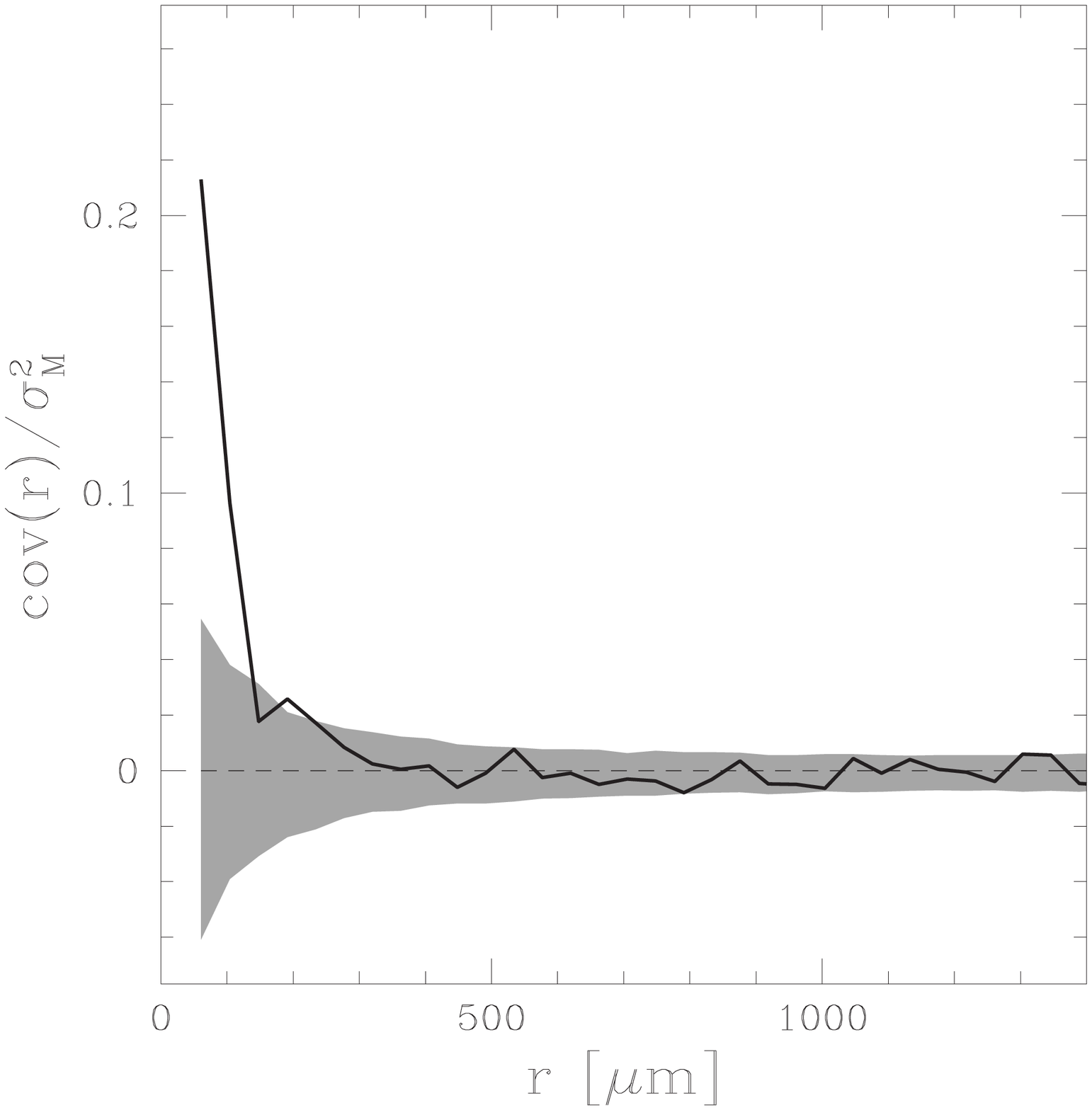,width=5cm}
\end{center}
\caption{The mark-weighted correlation functions from the holes in
the Fontainbleau sandstone. The pores' radii serve as marks. The $k_m$
being smaller than one indicates a depletion effect. The shaded areas again denote the range of
one-$\sigma$ fluctuations for randomized marks around the case of no
mark segregation.  The fluctuations were estimated from 200
reshufflings of the radii. \label{fig:kerscher_sand} }
\end{figure}

Now we turn to systems on  smaller scales.  Sandstone is an example of
a porous medium and  has extensively been investigated, mainly because
oil  was found  in the  pore network  of similar  stones. In  order to
extract the  oil from the  stone one  can try to  wash it out  using a
second liquid, e.g. water.  Therefore,  one tries to understand from a
theoretical point  of view, how  the microscopic geometry of  the pore
network determines  the macroscopic  properties of such  a multi-phase
flow. Especially  the topology and connectivity of  the microcaves and
tunnels prove  to be  crucial for the  flow properties  at macroscopic
scales.  Details are  given,  for instance,  in  the contributions  by
C.~Arns et  al., H.-J.~Vogel  et al. and  J.~Ohser in this  volume.  A
sensible physical model, therefore, in  the first place has to rely on
a thorough description of the pore pattern.
\\
One way to understand the pore network is to think of it as a union of
simple  geometrical  bodies.   Following \cite{kerscher_sok:direct},  one  can
identify distinct  pores together with  their position and  their pore
radius or extension.  This allows us to understand  the pore structure
in  terms of  a marked  point process,  where the  marks are  the pore
radii.
\\
In the following, we consider three-dimensional data taken from one of
the   Fontainbleau  sandstone   samples   through  synchrotron   X-ray
tomography.  These  data trace a  $4.52$ mm diameter  cylindrical core
extracted from a block with bulk porosity $\phi=13\%,$, where the bulk
porosity is  the volume  fraction occupied by  the pores.   A piece
with $2.91$mm
length (resulting  in a $46.7$ mm$^3$  volume) of the  core was imaged
and                    tomographically                   reconstructed
{}\cite{kerscher_flanery:thredimensional,kerscher_spanne:synchrotron,kerscher_arns:euler,kerscher_arns:parallel}.
Further details  of this sample  are presented in the  contribution by
C.~Arns et al. in this  volume.  Based on the reconstructed images the
positions of  pores and  their radii were  identified as  described in
{}\cite{kerscher_sok:direct}.  \\
In our results  for the mark correlation functions  a strong depletion
of   $k_m(r)$   and   $var(r)$   is   visible   for   $r<200\mu$m   in
Fig.~\ref{fig:kerscher_sand}.
\begin{figure}
\begin{center}
\epsfig{file=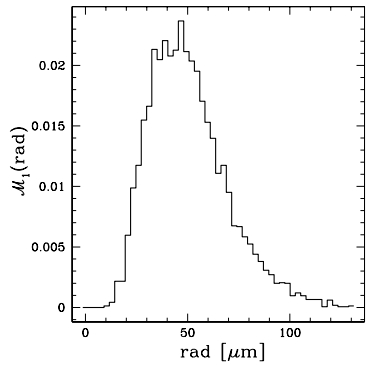,width=5cm}
\end{center}
\caption{The empirical one-point distribution ${\cal M}_1$ of the pore
sizes. \label{fig:kerscher_sand-hist} }
\end{figure}
This small-scale effect may be explained similarly to the Martian
craters: large pores are never found close to each others, since they
have to be separated by at least the sum of their radii. The histogram
of the pore radii in Fig.~\ref{fig:kerscher_sand-hist} shows that most of the
pores have radii smaller than 100$\mu$m, and consequently this effect
is confined to $r<200\mu$m.  In Sect.~\ref{sec:kerscher_boolean-depletion} we
discuss the Boolean depletion model which is based on this geometric
constraints and is able to produce such a reduction in the $k_m(r)$.
This purely geometric constraint also explains the reduced $var(r)$
and increased covariance $cov(r)$.
For separations larger than $200\mu$m there is no signal from the
covariance, but both $k_m(r)$ and $var(r)$ show a small increase out
to $\sim1000\mu$m. This indicate that pairs of pores out to these
separations tend to be larger in size and show slightly increased
fluctuations. However, this effect is small (of the order of  $1\%$)
and may be explained by the  definition of the holes, which
may lead to ``artificial small pores'' as ``bridges'' between larger
ones. This hypothesis has to be tested using different hole
definitions. In any case the main conclusion seems to be that apart from the
depletion effect at small scales there are no other mark correlations.
%%%%%%%%%%%%%%%%%%%%%%%
%       models
%%%%%%%%%%%%%%%%%%%%%%%
\section{Models for marked point processes}
\label{sec:kerscher_models}
Given the significant mark correlations found in various applications,
one may ask how these signals can be understood in terms of stochastic
models.   A  thorough  understanding  of course  requires  a  physical
modeling of the individual situation. There are, however, some generic
models, which we will focus on in the following:
in Sect.~\ref{sec:kerscher_boolean-depletion} we introduce the Boolean
depletion model, which is able to explain some of the features
observed in the distribution of craters and pores in sandstone.
Another generic model is the {\em random field model} where the marks
of the points stem from an independent random field
(Section~\ref{sec:kerscher_rfm}).  In Sect.~\ref{sec:kerscher_cox-rf-model} we
generalize the idea behind the random field model further in order to
get the {\em Cox random field model}, which allows for correlations
between the point set and the random field.
Other model classes and their applications are discussed by e.g.\
{}\cite{kerscher_diggle:statistical,kerscher_ogata:estimation,kerscher_cressie:statistics,kerscher_stoyan:fractals,kerscher_waelder:models}.

%%%
\subsection{The Boolean depletion model}
\label{sec:kerscher_boolean-depletion}

In our analysis of the Martian  craters and the holes in sandstone, we
found that for small separations only small craters, or small holes in
the sandstone, could be found. We interpreted this as a pure geometric
selection  effect.  The Boolean  depletion model  is able  to quantify
this effect, but also shows further interesting features.
\\
The  starting  point  is  the  Boolean model  of  overlapping  spheres
$B_R({\mathbf{x}})$ (see also the contributions by C.~Arns et al. and D.~Hug in
this volume as well  as \cite{kerscher_stoyan:lnp}).  For that, the spheres'
centers   ${\mathbf{x}}_i$   are    generated   randomly   and   independently,
i.e.  according to a  Poisson process  of number  density $\varrho_0$.
The radii $R$  of the spheres are then  chosen independently according
to a  distribution function  $F_0(R)$, i.e.  with  probability density
$f_0(R) =  \frac{\partial F_0(R)}{\partial R}$.  The  main idea behind
the depletion is to delete spheres which are covered by other spheres.
To make this  procedure unique we remove only  those spheres which are
completely covered by  a (notably larger) sphere\footnote{This process
can be thought  of as a dilution of the  original Poisson process, for
some   general    remarks   on   diluting    Poisson   processes   see
{}\cite{kerscher_stoyan:stochgeom}, p.~163. A  comparable model was considered
by~\cite{kerscher_stoyan:thinning}.}. The positions and radii of the remaining
spheres  define a  marked point  process.  Note,  that  this depletion
mechanism is  minimal in the sense  that a lot  of overlapping spheres
may  remain. This  Boolean depletion  model may  be considered  as the
low-density limit  of the  well-known Widom-Rowlinson model,  or (more
generally)  of  non-additive hard  sphere  mixtures  (see \cite{kerscher_widom:model,kerscher_mecke:additivity,kerscher_loewen:lnp}).
\\ 
The probability that a sphere of radius $R$ is not removed is then
given by
\begin{align}
\label{eq:kerscher_def-fnr}
f_{\rm nr}(R) 
& = \lim_{N,\Omega\rightarrow\infty} 
\prod_{i=1}^N \int_0^\infty {\rm d} R_i \ f_0(R_i) 
\left(1- \frac{4\pi}{3} \frac{(R_i-R)^3}{|\Omega|} \Theta(R_i-R) \right)\\
& = \exp\left( -\varrho_0 \omega_d  \int_0^\infty {\rm d} x \ f_0(R+x) x^d \right) ,  \nonumber
\end{align}
with  the  step function  $\Theta(x)=0$  for  $x<1$ and  $\Theta(x)=1$
otherwise, and the volume of the $d$-dimensional unit ball $\omega_d$
($\omega_1=2$,  $\omega_2=\pi$,   $\omega_3=4\pi/3$).   The  limit  in
Eq.~\eqref{eq:kerscher_def-fnr} is  performed by keeping $\varrho_0=N/|\Omega|$
constant, with  $N$ the initial  number of spheres and  $|\Omega|$ the
volume of the domain.
\\
The number density of the remaining spheres reads 
\begin{equation}\label{eq:kerscher_depletion-density}
\varrho = \varrho_0 \int_0^\infty {\rm d} R\ f_0(R) f_{\rm nr}(R) \;\;\;,
\end{equation}
where the one-point probability density ${\cal M}_1(R)$ that a sphere has radius $R$ is
given by
\begin{equation}
\label{eq:kerscher_M1-depletion}
{\cal M}_1(R) = f_0(R) f_{\rm nr}(R) \frac{\varrho_0}{\varrho} . 
\end{equation}
The probability  that one or both of
 the spheres $B_{R_1}({\mathbf{x}}_1)$ and $B_{R_2}({\mathbf{x}}_2)$
 are not removed is given by
\begin{equation}
f_{\rm nr}({\mathbf{x}}_1,R_1;{\mathbf{x}}_2,R_2) = 
\begin{cases}
0  &  \text{if } r < |R_2-R_1| ,\\
\exp\left(-\varrho_0 g_{\rm nr}({\mathbf{x}}_1,R_1;{\mathbf{x}}_2,R_2) \right) & \text{otherwise } , 
\end{cases}
\end{equation}
with  $B_{R<0}({\mathbf{x}})=\emptyset$ and  
\begin{equation}
g_{\rm nr}({\mathbf{x}}_1,R_1;{\mathbf{x}}_2,R_2) = \int_0^\infty {\rm d} x \; 
V(B_{x-R_1}({\mathbf{x}}_1)\cup B_{x-R_2}({\mathbf{x}}_2))\ f_0(x) .
\end{equation} 
At this  point we have to consider  the set union of  two spheres with
radii  $b_1\equiv  x-R_1$  and  $b_2\equiv x-R_2$,  respectively;  the
volume of this geometrical configuration can be calculated; in three
dimensions , e.g., we have:
\begin{align}
V(B_{b_1}(\vec{x}_1)\cup B_{b_2}(\vec{x}_2)) &= \frac{2\pi}{3}(b_1^3+b_2^3)\\ 
&\quad- \frac{2\pi}{ 3}\left( 
\frac{r^3}{8} - \frac{3}{4} r(b_1^2+b_2^2) - \frac{3}{8r}(b_2^2-b_2^2)^2 \right)  \notag
\end{align} 
for $|b_2-b_1|  \leq r  = |{\mathbf{x}}_2-{\mathbf{x}}_2|\leq b_1+b_2$.   Otherwise this
volume   reduces  either   to  the   volume  of   the   larger  sphere
($r<|b_2-b_1|$)  or to  the  sum of  both spherical  volumes ($r>b_1+b_2$).   \\
Similarly  as in  Eq.~\eqref{eq:kerscher_depletion-density}  the spatial two-point
density turns out to be
\begin{equation}
\label{eq:kerscher_rho2-depletion}
\varrho_2^{\cal S}({\mathbf{x}}_1,{\mathbf{x}}_2)  = 
\varrho_0^2 \int_0^\infty{\rm d} R_1 \int_0^\infty{\rm d} R_2 
f_0(R_1)f_0(R_2) f_{\rm nr}({\mathbf{x}}_1,R_1;{\mathbf{x}}_2,R_2) \;\;\;,
\end{equation}
such that the  conditional two-point mark density simply reads
\begin{equation} 
\label{eq:kerscher_M2-depletion}
{\cal M}_2(R_1,R_2|{\mathbf{x}}_1,{\mathbf{x}}_2) = f_0(R_1)f_0(R_2) f_{\rm nr}({\mathbf{x}}_1,R_1;{\mathbf{x}}_2,R_2) 
\frac{\varrho_0^2}{\varrho_2^{\cal S}({\mathbf{x}}_1,{\mathbf{x}}_2)} . 
\end{equation}
From this we can derive all of  the mark
correlation functions from Sect.~\ref{sec:kerscher_mark-segregation}.

%%%
\subsubsection{A bimodal distribution:}
\begin{figure}
\begin{center}
\epsfig{file=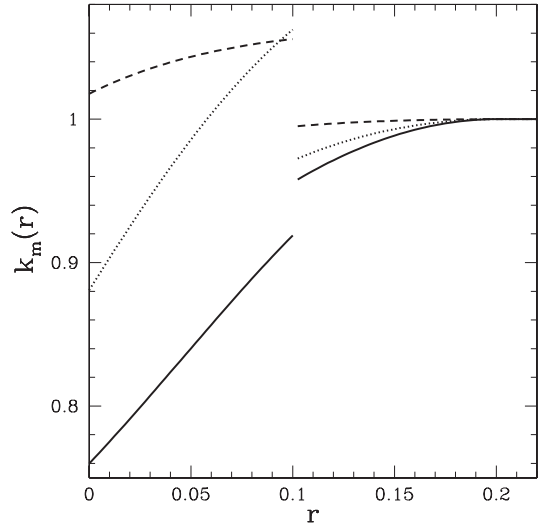,width=7cm}
\end{center}
\caption{ The $k_m(r)$ function for the Boolean depletion model with
parameters $R_1=0.05$, $R_2=0.15$, $\varrho=500$, and $\alpha=0.5$ (solid
line), $\alpha=0.3$ (dotted line), $\alpha=0.1$ (dashed line).
\label{fig:kerscher_depletion}}
\end{figure}

In order to get an analytically tractable model we adopt a bimodal
radius distribution  in the original Boolean model and start therefore with
\begin{equation} 
f_0(R) = \alpha_0 \delta(R-R_1) + (1-\alpha_0) \delta(R-R_2) \;\;, 
\end{equation}
where we assume that $R_1<R_2$.
Due to the depletion the number density $\varrho$ of the spheres as well as the
probability $\alpha$ to find the smaller radius $R_2$ at a given point are then
lowered; we get
\begin{align}
\alpha 
& = \alpha_0 \frac{{\rm e}^{-n}}{1-\alpha_0+\alpha_0{\rm e}^{-n}}\leq  \alpha_0 ,\\
\varrho 
& = \varrho_0\left( 1-\alpha_0 + \alpha_0{\rm e}^{-n} \right) = \varrho_0 \frac{1-\alpha_0}{1-\alpha} 
\end{align}
with $n        = \varrho (1-\alpha) \frac{4\pi}{3} (R_2-R_1)^3$. Altogether, the bimodal model can be parameterized in terms of the radii $R_1$ $R_2$, the
ratio $\alpha_0\in [0,1]$ and the density $\varrho_0\in{\mathbf{R}}^+$. The latter
two quantities, however are not observable from the final point process, therefore we convert them  into the  parameters $\alpha\in [0,1]$ and
$\varrho\in{\mathbf{R}}^+$, so that all other quantities can be expressed in
terms of these, for instance,
$\alpha_0  = \frac{\alpha}{\alpha + (1-\alpha){\rm e}^{-n}}  \geq
\alpha$, and
$\varrho_0    = \varrho \alpha {\rm e}^{n} + \varrho (1-\alpha)$,   
\\
From Eq.~\eqref{eq:kerscher_M1-depletion} we determine the mean mark, i.e.\ the
mean radius of the spheres
\begin{equation} 
\overline{m} = \overline{R} = \alpha R_1 + (1-\alpha) R_2 , 
\end{equation}
and from Eq.~\eqref{eq:kerscher_rho2-depletion} the spatial product density  
\begin{equation}
\varrho_2^{\cal S}(r) =\varrho^2 
\begin{cases}    
(1-\alpha)^2 R_2  +  \alpha^2 R_1 \exp\left(nI(x)\right)
%(1-\alpha)^2 + \alpha^2 \exp\left(nI(x)\right)
   & 0 \leq x <1  ,\\[1ex]
1+\alpha^2 \left[\exp\left(nI(x)\right)-1\right]
   &  1 \leq  x < 2 ,\\[1ex]
1  &  2 \leq x  ,
\end{cases} 
\end{equation}
with the normalized inter-sectional volume
$I(x)=1-\frac{3}{4}x+\frac{1}{16}x^3$ of two spheres and
$x=\frac{r}{|R_2-R_1|}$.
Finally, using Eq.~\eqref{eq:kerscher_M2-depletion} one can calculate the mark
correlation functions, e.g.
\begin{equation}
k_m(r) =
\begin{cases}
1-\alpha^2(1-\alpha) \frac{R_2-R_1}{\overline{R}} 
\frac{\exp\left(n I(x) \right)-\alpha^{-1}+1}
{(1-\alpha)^2+\alpha^2  \exp\left(n I(x) \right) }
%
%1-\alpha^2(1-\alpha) \frac{R_2-R_1}{\overline{R}} 
%\frac{\exp\left(n I(x) \right)-\alpha^{-1}+1}
%     {1+\alpha^2\left[\exp\left(n I(x) \right)-1\right] }   
& 0 \leq x <1, \\[1ex]
1-\alpha^2(1-\alpha) \frac{R_2-R_1}{\overline{R}} 
\frac{\exp\left(n I(x) \right)-1}
     {1+\alpha^2\left[\exp\left(n I(x) \right)-1\right] } 
&  1  \leq x < 2, \\[1ex]
1  &  2\leq x .
\end{cases}\label{eq:kerscher_km_depl}
\end{equation}
In Fig.~\ref{fig:kerscher_depletion} the $k_m(r)$ function from the Boolean
depletion model is shown. The model with the solid line illustrates
that a reduced $k_m(r)$ for small radii can be obtained by simply
removing smaller spheres. At least qualitatively this model is able to
explain the depletion effects we have seen both in the distribution
of Martian craters (Fig.~\ref{fig:kerscher_mars-markcorr}) and in the
distribution of pores in sandstone (Fig.~\ref{fig:kerscher_sand}). The jump at
$r=R_2-R_1$ is a relict of the strictly bimodal distribution with only
two radii. Fig.~\ref{fig:kerscher_depletion} also shows that the Boolean
depletion model is quite flexible, allowing for a $k_m(r)<1$, but also
$k_m(r)>1$ is possible.
\\
Without ignoring the considerable difference of this Boolean depletion
model to the pore size distribution in real sandstones (see
Figures~\ref{fig:kerscher_sandstone_eg}-\ref{fig:kerscher_sand}) one may still recognize
some interesting similarities: This simple model explains naturally a
decrease of $k_m(r)$ if the distribution of the radii is symmetric
($\alpha=1/2$). As visible in Figure~\ref{fig:kerscher_sand-hist} this is
approximately the case for the pore radii. Moreover, note that even
quantitative features are captured correctly indicating that the
decrease of  $k_m(r)$ visible in Figure~\ref{fig:kerscher_sand} is indeed due
to a depletion effect. For instance, the decrease starts at $r\approx
R_M$ where $R_M \approx 100 \mu m$ is the largest occurring radius (see
the histogram in Figure~\ref{fig:kerscher_sand-hist}) and the value of
$k_m(0)\approx 0.8$ at $r=0$ is in accordance with
Equation~\eqref{eq:kerscher_km_depl} assuming that $R_2-R_1\approx \overline{R}$
and the normalized density of pores $n\approx 1$ necessary for a
connected network. Of course a more detailed analysis is necessary
based on Eqs.~\eqref{eq:kerscher_M1-depletion} and ~\eqref{eq:kerscher_M2-depletion} and
the histogram shown in Figure~\ref{fig:kerscher_sand-hist}.

%%%
\subsection{The random field model}
\label{sec:kerscher_rfm}

The ``random-field model'' covers a class of models motivated
from fields such as geology (see, e.g.,\ {}\cite{kerscher_waelder:variograms}). The level
of the ground water, for instance, is thought of as a realization
of a random field which may be directly sampled at points (hopefully) independent
from the value of the field or which may influence the size of a tree
in a forest.\\
In general, a realization of the random field model is constructed from a
realization of a point process and a realization of a random field
$u({\mathbf{x}})$. The mark of  each object located at ${\mathbf{x}}_i$ traces the
accompanying random field via $m_i=u({\mathbf{x}}_i)$.  The crucial assumption
is that the point process is stochastically independent from the
random field.\\
We denote the  mean value of the homogeneous
random field by $\overline{u}={\mathbb{E}}[u({\mathbf{x}})]=\overline{u^1}$ and the
moments by $\overline{u^k}=\int{\rm d} u\ w(u) u^k$, with the one-point
probability density $w$ of the random field and ${\mathbb{E}}$ the expectation
over realizations of the random field. The product density of the
random field is $\rho_2^u(r)={\mathbb{E}}\big[u({\mathbf{x}}_1)u({\mathbf{x}}_2)\big]$ with
$r=|{\mathbf{x}}_1-{\mathbf{x}}_2|$. For a general discussion
of random field models, see {}\cite{kerscher_adler:randomfields}.
\\
In this model the one-point density of the marks is ${\cal M}_1(m)=w(m)$,
and $\overline{m^k}=\overline{u^k}$ etc.  The conditional mark density is
given by
\begin{equation}
{\cal M}_2(m_1,m_2|{\mathbf{x}}_1,{\mathbf{x}}_2) = 
{\mathbb{E}}\big[ \delta(m_1-u({\mathbf{x}}_1))\delta(m_2-u({\mathbf{x}}_2))\big],
\end{equation}
where $\delta$ is the Dirac delta distribution.  Clearly, this
expression is only well-defined under a suitable integral over the
marks.
With Eq.~\eqref{eq:kerscher_def-paverage} one obtains
\begin{equation}
\paverage{m_1}(r) =\overline{u} ,\quad
\paverage{m_1^2}(r) =\overline{u^2},\quad
\paverage{m_1 m_2}(r) = \rho_2^u(r) ,
\end{equation}
and the mark-correlation functions defined in
Sect.~\ref{sec:kerscher_mark-segregation} read
\begin{gather}
k_m(r) = 1,\
k_{mm}(r) = \rho_2^u(r)/\overline{u}^2,\
\gamma(r)=\overline{u^2}-\rho_2^u(r),\nonumber\\
\cov(r) = \rho_2^u(r)-\overline{u}^2,\
\var(r) = \overline{u^2} - \overline{u}^2 =\sigma_M^2.
\end{gather}
Therefore, there are some explicit predictions for the random field
model: an empirically determined $k_m$ significantly differing from
one not only indicates mark segregation, but also that the data is
incompatible with the random field model.  Looking at
Figure~\ref{fig:kerscher_ssrs-lum} we see immediately that the galaxy data are
not consistent with the random field model.
Similar tests based on the relation between $k_{mm}$ and the
mark-variogram $\gamma$ were investigated by
{}\cite{kerscher_waelder:variograms} and {}\cite{kerscher_schlather:characterization}.
The  failure of  the random  field  model to  describe the  luminosity
segregation in the galaxy  distribution allows the following plausible
physical  interpretation:   the  galaxies  do  not   merely  trace  an
independent  luminosity  field; rather  the  luminosities of  galaxies
depend on the clustering of the galaxies.
We shall try to account for  this with a better model in the following
section.

%%%
%%%%%%%%%%%%%%%%%%%%%%%%%%%%%%%%%%%%%%%%%%%%%%%%%%
%   the cox random field model
%%%%%%%%%%%%%%%%%%%%%%%%%%%%%%%%%%%%%%%%%%%%%%%%%%
\subsection{The Cox random field model}
\label{sec:kerscher_cox-rf-model}

In the  random field model,  the field was  only used to  generate the
points' marks.   In the Cox random  field model, on  the contrary, the
random  field determines  the spatial  distribution of  the  points as
well.  As  before, consider a  homogeneous and isotropic  random field
$u({\mathbf{x}})\ge0$.  The point process is constructed as a Cox-process (see
e.g.~\cite{kerscher_stoyan:stochgeom}).  The mean number  of points in a set
$B$ is given by the intensity measure
\begin{equation}
\label{eq:kerscher_cox-definition}
\Lambda(B) =\int_B{\rm d}{\mathbf{x}}\ a\ u({\mathbf{x}}) ,
\end{equation} 
where $a$ is a proportionality factor fixing the mean number density
$\varrho=a\overline{u}$. The (spatial) product  density of the point
distribution is
\begin{equation}
\varrho_2^{\cal S}({\mathbf{x}}_1,{\mathbf{x}}_2) = a^2\ \rho_2^u(r) 
= a^2\ \overline{u}^2 (1+\xi_2^u(r)),\label{eq:kerscher_xi_def}
\end{equation}
where again $\rho_2^u(r)$ denotes the product density of the random
field. $\xi_2^u$ is the normalized two-point cumulant of the random
field (see below). We will also need the $n$-point densities of the random field:
\begin{equation}
\rho_n^u({\mathbf{x}}_1,\ldots,{\mathbf{x}}_n)={\mathbb{E}}\big[u({\mathbf{x}}_1)\cdots u({\mathbf{x}}_1)\big].
\end{equation}
Like in  the random field model,  the marks trace the  field, but this
time rather  in a probabilistic way  than in a  deterministic one: the
mark $m_i$  on a galaxy located  at ${\mathbf{x}}_i$ is a  random variable with
the probability  density $p(m_i|u({\mathbf{x}}_i))$  depending on the  value of
the  field $u({\mathbf{x}}_i)$ at  ${\mathbf{x}}_i$. This  can be  used as  a stochastic
model  for the  genesis  of  galaxies depending  on  the local  matter
density.
\\ 
In order to calculate the conditional mark correlation functions we
define the conditional moments of the mark distribution given the
value $u$ of the random field:
\begin{equation}
\overline{m^k}(u) = \int{\rm d} m\ p(m|u)\ m^k.
\end{equation}
The spatial mark product-density is
\begin{equation}
\varrho_2^{SM}(({\mathbf{x}}_1,m_1),({\mathbf{x}}_2,m_2)) =
a^2{\mathbb{E}}\big[ p(m_1|u({\mathbf{x}}_1))p(m_2|u({\mathbf{x}}_2))\ u({\mathbf{x}}_1) u({\mathbf{x}}_2)\big].
\end{equation}
and with Eq.~\eqref{eq:kerscher_cond-mark-density}
\begin{equation}
{\cal M}_2(m_1,m_2|{\mathbf{x}}_1,{\mathbf{x}}_2) = 
\frac{1}{\rho_2^u(r)} 
{\mathbb{E}}\big[ p(m_1|u({\mathbf{x}}_1))p(m_2|u({\mathbf{x}}_2))\ u({\mathbf{x}}_1) u({\mathbf{x}}_2)\big] ,
\end{equation}
for  $\rho_2^u(r)\ne0$  and  zero  otherwise.   The  mark  correlation
functions can therefore be expressed in terms of weighted correlations
of the random field:
\begin{gather}
\paverage{m}(r) = \frac{1}{\rho_2^u(r)} {\mathbb{E}}\big[ \overline{m}(u({\mathbf{x}}_1))\ u({\mathbf{x}}_1) u({\mathbf{x}}_2)\big],\nonumber\\
\paverage{m^2}(r) = \frac{1}{\rho_2^u(r)} {\mathbb{E}}\Big[ \overline{m^2}(u({\mathbf{x}}_1))\ u({\mathbf{x}}_1) u({\mathbf{x}}_2)\Big],\\
\paverage{m_1 m_2}(r) = \frac{1}{\rho_2^u(r)} {\mathbb{E}}\big[ 
\overline{m}(u({\mathbf{x}}_1))\overline{m}(u({\mathbf{x}}_2))\ u({\mathbf{x}}_1) u({\mathbf{x}}_2)\big].\nonumber
\end{gather}

\subsubsection{A special choice for $p(m|u)$:}
To proceed further, we have to specify $p(m|u)$.  As a simple example
we choose $m_i$ equal to the value of the field $u({\mathbf{x}}_i)$ at the
point ${\mathbf{x}}_i$, such as in the random field model.  Thinking of the
random field as a mass density field and the mark of a galaxy
luminosity, that means that the galaxies trace the density field and
that their luminosities are directly proportional to the value of the
field.  With $p(m|u)=\delta(m-u)$ the conditional mark moments become
$\overline{m^k}(u)=u^k$.  The moments of the unconstrained mark
distribution read $\overline{m^k}=\overline{u^{k+1}}/\overline{u}$,
and the three basic pair averages are
\begin{gather}
\paverage{m_1}(r)= \frac{\rho_3^u({\mathbf{x}}_1,{\mathbf{x}}_1,{\mathbf{x}}_2)}{\rho_2^u(r)} ,\ 
\paverage{m_1^2}(r)= \frac{\rho_4^u({\mathbf{x}}_1,{\mathbf{x}}_1,{\mathbf{x}}_1,{\mathbf{x}}_2)}{\rho_2^u(r)} 
\nonumber\\
\paverage{m_1 m_2}(r)=\frac{\rho_4^u({\mathbf{x}}_1,{\mathbf{x}}_1,{\mathbf{x}}_2,{\mathbf{x}}_2)}{\rho_2^u(r)}  .
\end{gather}
Hence, the mark correlation functions defined in
Sect.~\ref{sec:kerscher_mark-segregation} are determined by the higher-order
correlations of the random field. With the Cox random field model we
go beyond the random field model, e.g.
\begin{equation}
\label{eq:kerscher_km-cox-model1}
k_m(r)=\frac{\paverage{m}(r)}{\overline{m}} = 
\frac{\overline{u}\ \rho_3^u({\mathbf{x}}_1,{\mathbf{x}}_1,{\mathbf{x}}_2)}{\overline{u^2}\ \rho_2^u(r)}
\end{equation}
is not equal to one any more.

\subsubsection{Hierarchical field correlations:}
\begin{figure}
\begin{center}
\epsfig{file=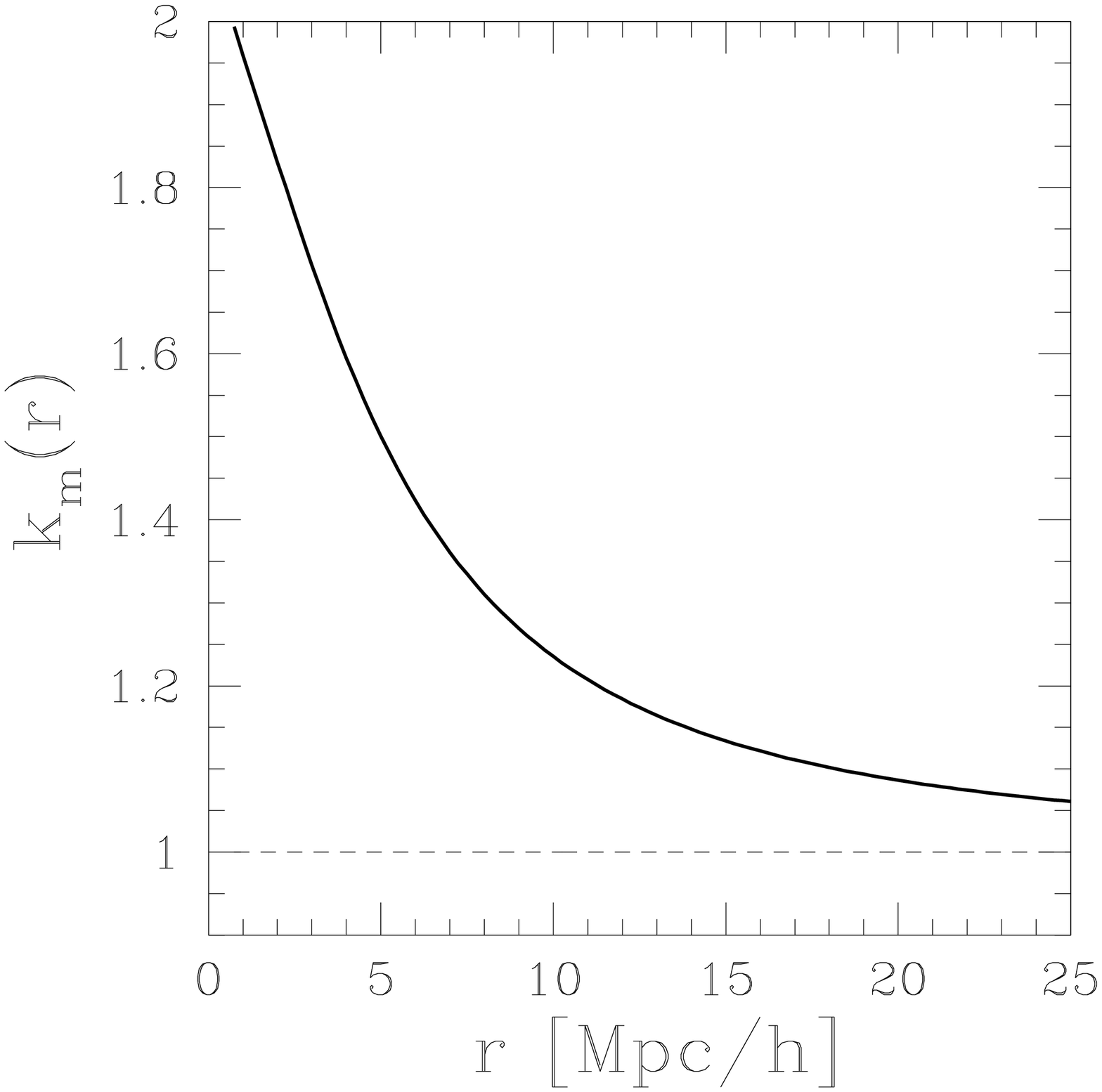,width=8cm}
\end{center}
\caption{\label{fig:kerscher_km-coxrf} The $k_m(r)$ function for the Cox random
field model according to Eq.~\eqref{eq:kerscher_km-cox-model2}. We use $Q=1$
and $\xi_2^u(r)=(5{\ifmmode{h^{-1}{\rm Mpc}}\else{$h^{-1}$Mpc}\fi}/r)^{1.7}$ truncated on small scales at
$\xi_2^u(r<0.1{\ifmmode{h^{-1}{\rm Mpc}}\else{$h^{-1}$Mpc}\fi})=\sigma_u^2/\overline{u}^2=\xi_2^u(0)\sim750$. }
\end{figure}

At this point, we have to specify the correlations of the random field
$u({\mathbf{x}})$.   The  simplest choice,  a  Gaussian  random  field, is  not
feasible        here,       since        a        number       density
(cp. Eq.~\ref{eq:kerscher_cox-definition}) has to be strictly positive, whereas
the Gaussian  model allows for  negative values. Instead, we  will use
the  hierarchical ansatz: we  first express  the two-  and three-point
correlations  in terms  of  normalized cumulants  $\xi_2$ and  $\xi_3$
(see,                                                            e.g.,\
{}\cite{kerscher_daley:introduction,kerscher_balian:I,kerscher_kerscher:constructing}),
\begin{multline}
\rho_2^u({\mathbf{x}}_1,{\mathbf{x}}_2) = \overline{u}^2\Big(1+
\xi_2^u({\mathbf{x}}_1,{\mathbf{x}}_2)\Big),\\
\rho_3^u({\mathbf{x}}_1,{\mathbf{x}}_2,{\mathbf{x}}_3)=\overline{u}^3\Big(1+
\xi_2^u({\mathbf{x}}_1,{\mathbf{x}}_2)+\xi_2^u({\mathbf{x}}_2,{\mathbf{x}}_3)+\xi_2^u({\mathbf{x}}_1,{\mathbf{x}}_3)+
\xi_3^u({\mathbf{x}}_1,{\mathbf{x}}_2,{\mathbf{x}}_3)
\Big).
\end{multline}
In order  to eliminate $\xi_3^u$  we use the hierarchical  ansatz (see
e.g.\ {}\cite{kerscher_peebles:lss}):
\begin{multline}
\xi_3^u({\mathbf{x}}_1,{\mathbf{x}}_2,{\mathbf{x}}_3)=Q\Big(
\xi_2^u({\mathbf{x}}_1,{\mathbf{x}}_2)\xi_2^u({\mathbf{x}}_2,{\mathbf{x}}_3)+
\xi_2^u({\mathbf{x}}_2,{\mathbf{x}}_3)\xi_2^u({\mathbf{x}}_1,{\mathbf{x}}_3)\\
+\xi_2^u({\mathbf{x}}_1,{\mathbf{x}}_2)\xi_2^u({\mathbf{x}}_1,{\mathbf{x}}_3)\Big).
\end{multline}
This ansatz is in reasonable agreement with data from the galaxy
distribution, provided $Q$ is of the order of unity
({}\cite{kerscher_szapudi:higherapm}).  Several choices for $\xi_2(r)$ and
$Q$ lead to well-defined Cox point process models based on the random
field $u({\mathbf{x}})$ {}\cite{kerscher_balian:I,kerscher_szapudi:higher}.
Now we can express $k_m(r)$ from Eq.~\eqref{eq:kerscher_km-cox-model1} entirely
in terms of the two-point correlation function $\xi_2^u(r)$ of the
random field:
\begin{equation}
\label{eq:kerscher_km-cox-model2}
k_m(r) = \frac{1+2\xi_2^u(r)+\xi_2^u(0)+Q\Big(\xi_2^u(r)^2+
2\xi_2^u(r)\xi_2^u(0)\Big)}  
{\Big(1+\xi_2^u(r)\Big)\Big(1+\xi_2^u(0)\Big)} ,
\end{equation}
where      we      made       use      of      the      fact      that
$\sigma_u^2=\overline{u^2}-\overline{u}^2=\overline{u}^2\xi_2^u(0)$.
Inserting typical parameters found  from the spatial clustering of the
galaxy distribution we  see from Fig.~\ref{fig:kerscher_km-coxrf} that
the Cox  random field  model allows us  to qualitatively  describe the
observed             luminosity             segregation             in
Fig.\ref{fig:kerscher_ssrs-lum}. But the  amplitude of $k_m$ predicted
by this  model is too  high. The Cox  random field model,  however, is
quite flexible  in allowing for  different choices for  $p(m|u)$; also
different models for the higher-order correlations of the random field
may     be     used,    e.g.\     a     log-normal    random     field
{}\cite{kerscher_coles:lognormal,kerscher_moller:log}.   Clearly  more
work  is  needed  to  turn  this  into viable  model  for  the  galaxy
distribution.

%%%
%%%%%%%%%%%%%%%%%%%%%%%
%      Conclusions 
%%%%%%%%%%%%%%%%%%%%%%%
\section{Conclusions}

Whenever objects are sampled together with their spatial positions and
some of their intrinsic properties, marked point processes are the
stochastic models for those data sets.  Combining the spatial information
and the objects' inner properties one can constrain their generation
mechanism and their interactions.
\\
Developing the framework of marked point processes further and
outlining some of their general notions is thus of interest for
physical applications.  Let us therefore look at mark correlations
again from both a statistical and a physical perspective.  We focused
on two kinds of dependencies.
\\
On the one hand, one can always ask, whether objects of different
types ``know'' from each other.  From a statistical point of view,
this is the question whether the marked point process consists of two
completely independent sub-point processes.  Physically, this
concerns the question whether the objects have been generated together
and whether they interact with each other. 
\\
On the other hand, it is often interesting to know whether the spatial
distribution of the objects  changes with their inner properties.  For
the  statistician,  this translates  into  the  question whether  mark
segregation  or  mark-independent   clustering  is  present.  For  the
physicist such  a dependency is  interesting since one can  learn from
them whether  and how  the interactions distinguish  between different
object classes or  whether the formation of the  objects' mark depends
on the environment.
\\
We discussed statistics capable of probing to which extent mark
correlations are present in a given data set, and showed how to assess
the statistical significance.  Applying our statistics to real data,
we could demonstrate, that the clustering of galaxies depends on their
luminosities.  Large scale correlations of the orientations of dark
matter halos were found.  Using the Mars data we could validate a
picture of crater generation on the Martian surface: mainly, the local
geological setting determines the crater type. We also could show that
the sizes of pores in sandstone are correlated.
\\
In order to understand empirical data sets in detail, we need models
to compare to. As generic models the Boolean depletion model, the
random field model and its extension, the Cox random field models are
of interest.
\\
Further application of the mark correlations properties may  inspire the
development of  further models.  It seems therefore  that marked point
processes could spark  interesting interactions between physicists and
mathematicians.   Certainly,  the   distributions  of  physicists  and
mathematicians  in  coffee breaks  at  the  Wuppertal conference  were
clustered, each.  But could one  observe positive cross-correlations?
Using mark  correlations we argue, that,  even more, there  is lots of
space for positive interactions.\ldots.

\subsubsection*{Acknowledgments}
We would like to thank Andreas Faltenbacher, Stefan Gott\-l\"o\-ber and
Volker M\"uller for allowing us to present some results from the
orientation analysis of the dark matter halos (Sect.~\ref{sec:kerscher_halos}).
For providing the sandstone data (Sect.~\ref{sec:kerscher_sandstone}) and
discussion we thank Mark Knackstedt.
Herbert Wagner provided constant support and encouragement, especially
we would like to thank him for introducing us to the concepts of
geometric algebra as used in the Appendix.

\section*{Appendix: Completeness of mark correlation functions}
\label{sec:kerscher_app}

In  order to  form   versatile test   functions for   describing  mark
segregation effects,  we  integrated the conditional  mark probability
density ${\cal M}_2(m_1,m_2|r)$ twice  in mark space thereby weighting with
a       function      of       the      marks   $f(m_1,m_2)$      (see
Eq.~\ref{eq:kerscher_def-paverage}). Such  a pair-averaging  reduces the  full
information present in ${\cal M}_2(m_1,m_2|r)$.  So one may ask, whether or
in  which sense  the mark  correlation  functions  give  a complete
picture of the present two-point mark correlations.\\
For scalar   marks $m_i$  this  task is  trivial.  With  a  polynomial
weighting     function     $f(m_1,m_2)\sim    m_1^{n_1}     m_2^{n_2}$
($n_1,n_2=0,1,..$)  we  consider moments of ${\cal M}_2(m_1,m_2|r)$, hence,
we can be complete  only up to a  given polynomial order in the  marks
$m_1$ and $m_2$.
At first  order there is  only the mean $\paverage{m}(r)$.   At second
order  we have $\paverage{m^2}(r)$  and $\paverage{m_1  m_2}(r)$.  All
the       mark      correlation      functions       discussed      in
Sect.~\ref{sec:kerscher_mark-segregation} can be  constructed from these three
pair  averages\footnote{This completeness  of  $\paverage{m^2}(r)$ and
$\paverage{m_1  m_2}(r)$ at  the  two-point level,  however, does  not
imply that  one should not  consider linear combinations of  them. For
instance,  it  may  well  be   the  case,  that  only  certain  linear
combinations yield significant results.}.  Higher-order moments of the
marks involve more and more cross-terms.
\\
For vector-valued  marks, however,  it is not  obvious that  the test
quantities  proposed  in  Sect.~\ref{sec:kerscher_mark-segregation} trace  all
possible  correlations between  the  vectors up  to  third order.  To
settle  this case  we  have  to consider  the  framework of  geometric
algebra,  also called  Clifford algebra.   A detailed  introduction to
geometric   algebra  is   given   in  {}\cite{kerscher_hestens:new},   shorter
introductions are {}\cite{kerscher_gull:imaginary,kerscher_lasenby:unified}.
In geometric  algebra one  assigns  a unique meaning to  the geometric
product (or   Clifford product) of   quantities like vectors, directed
areas, directed volumes,  etc.  The geometric  product ${\mathbf{a}}{\mathbf{b}}$ of two
vectors  ${\mathbf{a}}$ and ${\mathbf{b}}$  splits into its  symmetric and antisymmetric
part
\begin{equation}
{\mathbf{a}}{\mathbf{b}} = {\mathbf{a}}\cdot{\mathbf{b}} + {\mathbf{a}}\wedge{\mathbf{b}}.
\end{equation}
Here  ${\mathbf{a}}\cdot{\mathbf{b}}$ denotes  the   usual   scalar product; in    three
dimensions, the wedge product ${\mathbf{a}}\wedge{\mathbf{b}}$ is closely related to the
cross product between  these two vectors.  However,  ${\mathbf{a}}\wedge{\mathbf{b}}$ is
not  a vector like  ${\mathbf{a}}\times{\mathbf{b}}$, but a bivector  -- a directed area.
Higher products of vectors can be simplified according to the rules of
geometric algebra (for details see {}\cite{kerscher_hestens:new}).
\\
Let us consider the situation  where objects  situated at ${\mathbf{x}}_1$  and
${\mathbf{x}}_2$ bear vector  marks ${\mathbf{l}}_1$ and  ${\mathbf{l}}_2$, respectively, and let
the  normalized distance vector  be $\hat{\mathbf{r}}=({\mathbf{x}}_1-{\mathbf{x}}_2)/r$.  Note, that
$\hat{\mathbf{r}}$ is  not a mark at all,  rather it can  be thought of  as another
vector which may be useful for constructing mark correlation functions.
\\
  For many applications it is   reasonable to assume isotropy in  mark
space, i.e.  all of the mark  correlation functions are invariant
under common  rotations of the  marks. For galaxies,  e.g., there does
not seem to be an a priori preferred direction for their orientation. In more detail we have then
\begin{align}
\begin{split}
{\cal M}_1 ({\mathbf{l}}) &= {\cal M}_1 (R{\mathbf{l}})= {\cal M}_1 (|{\mathbf{l}}|) \;\;\;,\\
{\cal M}_2 ({\mathbf{l}}_1,{\mathbf{l}}_2|r)&= {\cal M}_2(R{\mathbf{l}}_1,R{\mathbf{l}}_2|r)\;\;\;,\notag
\end{split}
\end{align}
and so on, where  $R$ is an arbitrary   rotation in mark space.   This
means that the mark correlation functions  depend only on rotationally
invariant    combinations   of  the  vector  marks.    Therefore, only
rotationally invariant combinations  of vectors are  sensible building
blocks for  weighting functions.  We   thus can restrict ourselves  to
scalar  weighting  functions, which result  in coordinate-independent
vector-mark correlation functions.
%
%In  a more  general setting one  might  be  interested in  $k$-vector
%valued   mark correlation   functions,  but   condensed   into  scalar
%information.
\\   Again  we  proceed   by  considering   mixed  moments   as  basic
combinations.   We  restrict  ourselves  to  scalar  quantities  being
polynomial  in the  vector components.   One may  also discuss
moments in a  broader sense allowing for vector  moduli. In this wider
sense,  for example,  $|{\mathbf{l}}_1|$  or $|{\mathbf{l}}_1  \times  ({\mathbf{l}}_1\times\hat
{\mathbf{r}})|$  would be  allowed. We  do  not consider  such quantities  here,
because they are not polynomial in the vector components. Their squares
anyway appear at higher orders.  Furthermore,
it turns out that the  characterization we will provide depends on the
embedding dimension. The first-  and second-order moments are identical
in two  and three  dimensions, but  at the third  order they  start to
differ.
\begin{enumerate}
\item In the strict sense of scalar quantities being linear in the
vector components there are no first-order moments for vectors. 
\item   At  second   order  we   encounter  the   following  products:
${\mathbf{l}}_1{\mathbf{l}}_1$,     $\hat{\mathbf{r}}\hat{\mathbf{r}}$,
${\mathbf{l}}_1{\mathbf{l}}_2$,       $\hat{\mathbf{r}}{\mathbf{l}}_1$.
Note,     that,     e.g.,     ${\mathbf{l}}_1\hat{\mathbf{r}}$     and
${\mathbf{l}}_2\hat{\mathbf{r}}$ do not make any difference as regards
the  mark correlation  functions, since  the pair  averages implicitly
render  the  indices  symmetric;  moreover, although  the  geometrical
product  is non-commutative, ${\mathbf{l}}_1\wedge  \hat {\mathbf{r}}$
and ${\mathbf{l}}_1\wedge \hat {\mathbf{r}}$  do not lead to different
mark           correlation           functions.           Furthermore,
$\hat{\mathbf{r}}\hat{\mathbf{r}}=1$.
${\mathbf{l}}_1{\mathbf{l}}_1={\mathbf{l}}_1\cdot{\mathbf{l}}_1=l_1^2$
provides  us with higher  moments of  the modulus  of the  vectors. To
investigate these kinds of  correlations already scalar marks would be
sufficient.   New information is  encoded in  the other  products.  \\
Consider
${\mathbf{l}}_1{\mathbf{l}}_2={\mathbf{l}}_1\cdot{\mathbf{l}}_2+{\mathbf{l}}_1\wedge{\mathbf{l}}_2$.
The  symmetric part  ${\mathbf{l}}_1\cdot{\mathbf{l}}_2$ is  clearly a
scalar     and     defines     the     alignment     ${\cal     A}(r)$
(Eq.~\ref{eq:kerscher_align}).        The      antisymmetric      part
${\mathbf{l}}_1\wedge{\mathbf{l}}_2$ is  a bivector. Its  -- unique --
modulus        (see        again       {}\cite{kerscher_hestens:new}),
$|{\mathbf{l}}_1\wedge{\mathbf{l}}_2|=\sqrt{l_1^2l_2^2-({\mathbf{l}}_1\cdot{\mathbf{l}}_2)^2}$,
may  be  useful,  but  is   no  longer  a  polynomial  in  the  vector
components.  $|{\mathbf{l}}_1\wedge{\mathbf{l}}_2|^2$  appears at  the
fourth   order.  In   a  completely   analogous  way   we   can  treat
${\mathbf{l}}_1\hat{\mathbf{r}}={\mathbf{l}}_1\cdot\hat{\mathbf{r}}+{\mathbf{l}}_1\wedge\hat{\mathbf{r}}$.
The   symmetric  part   ${\mathbf{l}}_1\cdot\hat{\mathbf{r}}$  defines
${\cal F}(r)$.   Hence at second order, the  only possible vector-mark
correlation functions are ${\cal A}(r)$ and ${\cal F}(r)$.
\item At third order we have to consider products of three vectors. In
general the product of three vectors ${\mathbf{a}},{\mathbf{b}},{\mathbf{c}}$ splits into 
\begin{equation}
{\mathbf{a}}{\mathbf{b}}{\mathbf{c}} = {\mathbf{a}}({\mathbf{b}}\cdot{\mathbf{c}}) + ({\mathbf{a}}\cdot{\mathbf{b}}){\mathbf{c}} - ({\mathbf{a}}\cdot{\mathbf{c}}){\mathbf{b}} 
+{\mathbf{a}}\wedge({\mathbf{b}}\wedge{\mathbf{c}}) .
\end{equation}
i.e.,  a  vector  (consisting  of   the  three  first  terms),  and  a
pseudo-scalar,   a   directed   volume.    In  two   dimensions   the
pseudo-scalar ${\mathbf{a}}\wedge({\mathbf{b}}\wedge{\mathbf{c}})$ vanishes.
 \\ Now  we have to  form all possible  products of the  three vectors
${\mathbf{l}}_1,{\mathbf{l}}_2,\hat{\mathbf{r}}$ and to derive  scalars.  In three dimensions the
only       new       combination       is      the       pseudo-scalar
${\mathbf{l}}_1\wedge({\mathbf{l}}_2\wedge\hat{\mathbf{r}})$    giving   the    oriented   volume
${\mathbf{l}}_1\cdot({\mathbf{l}}_2\times\hat{\mathbf{r}})$. Unfortunately,  this oriented volume
averages  out  to  zero.  Thus,  in  a  strict  sense,  there  are  no
interesting third-order  quantities. Closely related,  however, is the
modulus   of   the   pseudoscalar   $|{\mathbf{l}}_1\cdot({\mathbf{l}}_2\times\hat{\mathbf{r}})|$
proportional  to our  ${\cal P}(r)$.   This expression  is invariant  under
permutations of the vectors.
\item 
At third order and in  two dimensions all of the relevant combinations
are   products   of  first-   and   second-order  combinations;   no
specifically new combination appears.  This is different from the case
of three  dimensions, where at  third order an entirely  new geometric
object,  the pseudo-scalar  ${\mathbf{l}}_1\wedge({\mathbf{l}}_2\wedge\hat{\mathbf{r}})$  can be
constructed.
There  is  a  general  scheme  behind  this  argument:  since  in  $d$
dimensions any geometrical product  of more than $d$ vectors vanishes,
all relevant  combinations of  vectors at orders  higher than  $d$ are
essentially products of combinations of lower-order factors.
\end{enumerate}

%%%%
%%%%%

%\bibliographystyle{springer}
%\bibliography{wuppertal}
%\bibliography{my}

\end{document}